\documentclass[%
reprint,
superscriptaddress,
 amsmath,amssymb,
 aps,
]{revtex4-2}

\usepackage{graphicx}
\usepackage{dcolumn}
\usepackage{bm}
\usepackage{physics}
\usepackage{booktabs}
\usepackage{subfig}
\usepackage[ruled]{algorithm2e}
\usepackage{multirow}

\usepackage{caption}
\usepackage[normalem]{ulem} 
\usepackage{amsmath}
\usepackage{xcolor}         
\usepackage{soul}           
\usepackage{changes}        

\definecolor{added}{RGB}{0,100,0}      
\definecolor{deleted}{RGB}{150,0,0}    
\definecolor{comment}{RGB}{0,0,150}    

\preprint{APS/123-QED}

\begin{document}

\title{Neural Network Backflow with Low-Rank Multi-Determinant Updates}

\author{Tianshu Huang}
\email{tianshu.huang@yale.edu}
\affiliation{Energy Sciences Institute, Yale University, West Haven, Connecticut 06516, USA}
\affiliation{Department of Applied Physics, Yale University, New Haven, Connecticut 06520, USA}

\author{Xiaowei Ou}
\email{xiaowei.ou@yale.edu}
\affiliation{Energy Sciences Institute, Yale University, West Haven, Connecticut 06516, USA}
\affiliation{Department of Physics, Yale University, New Haven, Connecticut 06520, USA}

\author{Vidvuds Ozoli\c{n}\v{s}}%
\email{vidvuds.ozolins@yale.edu}
\affiliation{Energy Sciences Institute, Yale University, West Haven, Connecticut 06516, USA}
\affiliation{Department of Applied Physics, Yale University, New Haven, Connecticut 06520, USA}

\date{\today}

\begin{abstract}
Simulating strongly correlated fermions remains a long-standing challenge due to the exponential complexity of the Hilbert space and the intricate sign structure of many-body wavefunctions. We introduce a variational framework centered on a neural network backflow transformation that combines deep learning with variational Monte Carlo. The proposed ansatz employs a multi-determinant expansion with low-rank shifts to capture non-local correlations and complex sign structures. Applied to the two-dimensional Hubbard model at both half-filling and $1/8$ doping, the method achieves energies within $0.45\%$ of auxiliary-field quantum Monte Carlo at half-filling and captures intertwined charge- and spin-density stripe patterns at $1/8$ doping. These results demonstrate the potential of this framework as a scalable and interpretable approach to variational simulations of strongly correlated fermionic systems.
\end{abstract}

\maketitle

\section{Introduction}
Solving many-body fermionic systems remains one of the most formidable challenges in computational physics, primarily due to the exponential growth of the Hilbert space and the emergence of non-trivial wavefunction sign structures. These complexities often render traditional numerical methods either computationally prohibitive or systematically biased. To address this, many numerical methods are under active development, including the density matrix renormalization group (DMRG) \cite{schollwock2005density,stoudenmire2012studying}, projected entangled pair states (PEPS) \cite{vidal2007classical,cirac2021matrix,kraus2010fermionic}, and various quantum Monte Carlo (QMC) \cite{zhang1997constrained,zhang2003quantum,zhang201315,becca2017quantum,sugiyama1986auxiliary} techniques. Despite their successes, these methods face inherent limitations: tensor networks like DMRG and PEPS are constrained by high entanglement in two dimensions, while quantum Monte Carlo remains hampered by the fermionic sign problem, particularly in the computationally challenging doped regime.

In recent years, neural quantum states (NQS) \cite{carleo2017solving} have emerged as a transformative paradigm for addressing these gaps. By utilizing neural networks as flexible variational wavefunction ansatzes, NQS methods have demonstrated a remarkable ability to capture intricate many-body correlations \cite{wu2024variational, schmitt2022quantum, nys2024ab, chen2024empowering, rende2024simple, ou2025improving, viteritti2025transformer}. For fermionic systems, the application of NQS typically adapts the Slater determinant formalism to satisfy the requisite antisymmetry. To incorporate higher-order correlations, Jastrow factors, backflow transformations, and their variants are introduced \cite{nomura2017restricted, choo2020fermionic, pfau2020ab, hermann2020deep, liu2024neural, luo2019backflow, robledo2022fermionic, liu2024unifying, chen2025neural, gu2025solving}. In this framework, the neural network performs a backflow transformation that maps bare particle coordinates into a set of collective, quasiparticle features. The neural network backflow transformation has consistently provided some of the most accurate variational energies to date, as it is uniquely capable of deforming the nodal surfaces of the wavefunction to converge toward the true physical ground state.

However, the expressive power of deep neural network backflow transformations comes at a high cost: models with a large number of parameters are computationally expensive to store and involve intensive evaluation times, often limiting their scalability for larger systems. One established path to enhancing performance without simply scaling the network depth is the adoption of a multi-determinant wavefunction format. By representing the state as a linear combination of several Slater determinants, the model can capture different sectors of the Hilbert space more effectively. When calculating neural network backflow coefficients, one could directly generate unique backflow matrices for each determinant by modifying the number of network outputs \cite{gu2025solving}. Meanwhile, the parameter count blows up linearly with the number of determinants, quickly leading to memory bottlenecks. Alternatively, one can utilize the physical symmetries of the lattice to generate a multi-determinant state by evaluating a single backflow network on various symmetry-transformed versions of the electron configuration \cite{sharma2025comparing, loehr2025enhancing, romero2025spectroscopy, rodriguez2013multireference, nomura2021helping, reh2023optimizing}. While this approach is extremely parameter-efficient for reusing the same weights for all determinants while enforcing physical symmetries, it introduces additional computational cost for calculating the character of irreducible representations of the symmetry operations and the network evaluation of all symmetry-equivalent configurations.

We present matrix shifts as an alternative approach for generating multi-determinant expansions. To maintain computational efficiency, these determinant-specific shifts are implemented in a low-rank factorized form. By decomposing the updates into low-rank products rather than full-rank matrices, this architecture enables a multi-determinant ansatz while avoiding the over-parameterization and prohibitive scaling typically associated with traditional methods. This property ensures that both the parameter count and the cost of determinant updates remain numerically tractable.

\begin{figure*}
    \includegraphics[scale=0.5]{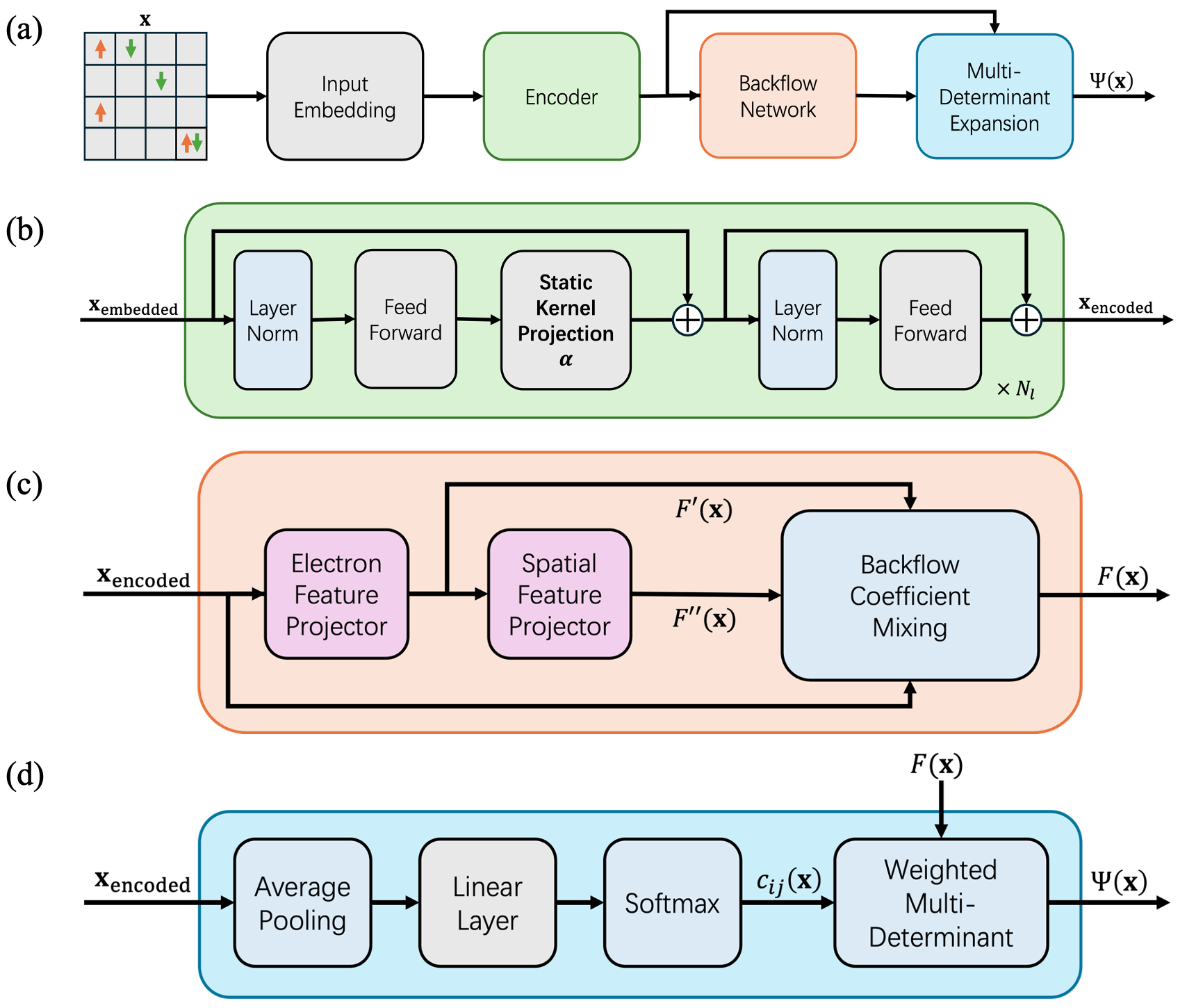}
    \centering
    \caption{Schematic of the neural network architecture. (a) The overall workflow, where the lattice configuration $\mathbf{x}$ is processed through an input embedding and encoder to feed both the backflow network and the determinant expansion. (b) The Encoder architecture, featuring $N_l$ layers of layer normalization, feed-forward blocks, and a static kernel projection $\alpha$ with residual connections to produce the encoded features $\mathbf{x}_{\text{encoded}}$. (c) The Backflow Network, which utilizes an electron feature projector and a spatial feature projector to generate mixed backflow coefficients $F(\mathbf{x})$. (d) The Multi-Determinant Expansion block, where encoded features are spatially averaged and passed through a linear layer with a Softmax activation to produce configuration-dependent weights $c_{ij}(\mathbf{x})$. These weights are combined with the backflow-transformed features to evaluate the final weighted multi-determinant wavefunction $\Psi(\mathbf{x})$.
    }
    \label{fig:network_architectures}
\end{figure*}

\section{Model and Method}

\subsection{The 2D Hubbard Hamiltonian}
We focus on the standard Hubbard Hamiltonian, which considers only nearest-neighbor hopping and on-site Coulomb interactions. The Hamiltonian is expressed as
\begin{equation}
\label{eq:Hubbard_Hamiltonian}
\hat{H} = -t \sum_{\sigma=\uparrow,\downarrow} \sum_{<i,j>}^{N_\text{site}} (\hat{c}^{\dagger}_{i\sigma}\hat{c}_{j\sigma} + \hat{c}^{\dagger}_{j\sigma}\hat{c}_{i\sigma}) + U\sum_{j}^{N_\text{site}}\hat{n}_{j\uparrow}\hat{n}_{j\downarrow},
\end{equation}
where $N_\text{site}$ is the total number of lattice sites. The operators $\hat{c}^{\dagger}_{i\sigma}$ and $\hat{c}_{i\sigma}$ denote the creation and annihilation of an electron with spin $\sigma$ at site $i$, respectively. The notation $\langle i,j \rangle$ indicates that hopping is restricted to nearest-neighbor pairs with an associated hopping amplitude $t$. The occupation number operator is given by $\hat{n}_{j\sigma} = \hat{c}^{\dagger}_{j\sigma}\hat{c}_{j\sigma}$, and $U$ represents the energy penalty for two electrons of opposite spin occupying the same site. The first term in Eq.~(\ref{eq:Hubbard_Hamiltonian}) represents the kinetic energy of the system, while the second term accounts for the on-site Coulomb repulsion. 

\subsection{Variational Monte Carlo and Optimization}
To determine the ground state of the Hubbard Hamiltonian, we optimize the variational parameters $\theta$ of the neural backflow ansatz using the Variational Monte Carlo (VMC) method. The primary objective is to minimize the energy expectation value, which serves as the loss function:
\begin{equation}
    E(\theta) = \frac{\langle \Psi_\theta | \hat{H} | \Psi_\theta \rangle}{\langle \Psi_\theta | \Psi_\theta \rangle} = \mathbb{E}_{\mathbf{x} \sim |\Psi\theta(\mathbf{x})|^2} [E_L(\mathbf{x})]
\end{equation}
where $E_L(\mathbf{x}) = \sum_{\mathbf{x}'} H_{\mathbf{x},\mathbf{x}'} \frac{\Psi_\theta(\mathbf{x}')}{\Psi_\theta(\mathbf{x})}$ is the local energy. We estimate this quantity and its gradients by sampling the configuration space according to the probability distribution $|\Psi_\theta(\mathbf{x})|^2$. 

To update the parameters $\theta$, we utilize the Subsampled Projected-Increment Natural Gradient Descent (SPRING) method \cite{goldshlager2024kaczmarz}. While the standard stochastic reconfiguration (SR) method \cite{sorella2005wave} identifies an update by minimizing the norm of the difference between the variational parameter shifts and the local energy gradient, SPRING also prioritizes stability by minimizing the deviation from the previous parameter update.

\subsection{Neural network ansatz}
We solve the ground-state properties of the two-dimensional Hubbard model using a variational Monte Carlo (VMC) framework enhanced by a deep neural network wavefunction ansatz. The central challenge in simulating the doped Hubbard regime lies in the intricate nodal structure of the fermionic wavefunction, which traditional trial states often fail to capture with sufficient flexibility. To address this, we introduce a hierarchical architecture that combines a deep neural network backflow transformation with a low-rank factorized multi-determinant expansion. Fig.~\ref{fig:network_architectures} presents our neural network architecture design. In the following subsections, we detail the components of this architecture.

\subsubsection{Input embedding}
The network architecture begins with an input embedding block that converts the input configuration $\mathbf{x}$, a 2D grid representing site occupations and spin states, into a format the model can ingest. Inspired by the vision transformer architecture \cite{dosovitskiy2020image}, we first split the lattice grid into $N_\text{patch}$ fixed-size patches of shape $p \times p$, each capturing a subset of the spatial occupation and spin information. Following the partitioning, each patch's spatial information is transformed into a single row vector, where the electron occupations are represented in a one-hot encoding format, and these rows are stacked to form a structured token matrix. The token matrix then passes through a linear layer that performs a trainable projection. This layer maps the discrete, sparse information of the electron occupations into a high-dimensional latent space. The resulting output is an embedded vector $\mathbf{x}_\text{embedded} \in \mathbb{R}^{N_\text{patch} \times d}$, where $d$ is the embedding dimension. 

\subsubsection{Static kernel encoder}
The embedded variable is fed into an encoder featuring a residual-based structure designed for efficient feature extraction. Each of the $N_l$ consecutive layers consists of two primary sub-blocks, both of which utilize layer normalization to ensure training stability. The first sub-block processes the input through a feed-forward network, followed by a projection from a static kernel $\alpha$. As an input-independent learnable parameter matrix, $\alpha$ enables the model to enforce specific spatial correlations directly within the mapping, a process analogous to a factored attention mechanism without translation-invariant modulation \cite{viteritti2025transformer}. A secondary feed-forward sub-block provides further feature refinement. Both stages incorporate residual connections ($\oplus$), which facilitate efficient gradient flow and allow the network to learn incremental updates to the data representation. The encoder finally outputs the processed state $\mathbf{x}_{\text{encoded}} \in \mathbb{R}^{N_\text{site} \times d}$.

\subsubsection{Neural network backflow transformation}
Neural network backflow transformation generalizes the traditional Feynman-Cohen backflow \cite{feynman1956energy} by using deep learning architectures to map bare electronic coordinates to a high-dimensional space \cite{luo2019backflow}. The backflow-transformed wavefunction has the form
\begin{equation}
\label{bf_wf}
\Psi_{BF}{(\mathbf{x})} = \det [M + F(\mathbf{x})],
\end{equation}
where $M$ is the single-particle orbital matrix, and $F(\mathbf{x})$ is the neural network backflow coefficient matrix calculated from the given electronic configuration $\mathbf{x}$. Both $M$ and $F(\mathbf{x})$ has the same shape $\mathbb{R}^{2N_\text{site} \times N_\text{electron}}$, where $N_\text{electron}$ is the number of electrons in the configuration. We stack the two spin sectors into one matrix, hence the shape $2N_\text{site}$ along the row dimension. When calculating the determinant, we select the electrons' corresponding occupied rows from $M + F(\mathbf{x})$ to obtain the matrix with proper shape $\mathbb{R}^{N_\text{electron} \times N_\text{electron}}$.

Instead of using a single linear layer to merely transform the output backflow matrix to the proper shape, we emphasize the expressive power of the backflow network by constructing a deep separable bilinear architecture. The latent vector $\mathbf{x}_{\text{encoded}}$ is first passed through an electron feature projector followed by a spatial feature projector. Each projector is a three-layer feed-forward network with Gaussian Error Linear Unit (GELU) \cite{hendrycks2016gaussian} activation functions in between. These two modules decouple and transform the particle-specific and coordinate-specific information into intermediate representations, denoted as $F'(\mathbf{x}) \in \mathbb{R}^{N_\text{site} \times N_\text{electron}}$ and $F''(\mathbf{x}) \in \mathbb{R}^{2N_\text{site} \times N_\text{electron}}$. Both the intermediate feature maps and the original encoded input are fed into a central backflow coefficient mixing module. This final stage acts as a fusion layer, combining the transformed spatial and electron features with a residual path from the initial encoding. This ensures that the final output $F(\mathbf{x})$ preserves global context while being conditioned on the specific local features extracted by the projectors. The final backflow matrix is expressed as 
\begin{equation}
\label{eq:multi_dets}
    F(\mathbf{x}) = F''(\mathbf{x}) \odot \left( J + W_1 \mathbf{x}_{\text{encoded}} Y_1 + W_2 F'(\mathbf{x}) Y_2 \right),
\end{equation}
where $\odot$ represents element-wise multiplications; $J$ is the matrix whose entries are all ones; $W_i$ and $Y_i$ are learnable parameter matrices that transform $\mathbf{x}_{\text{encoded}}$ and $F'(\mathbf{x})$ to the same shape as $F''(\mathbf{x})$.

\subsection{Multi-determinant wavefunction with low-rank shifts}
Our multi-determinant wavefunction with low-rank shifts takes the form:
\begin{equation}
\Psi(\mathbf{x}) = \sum_{i,j} c_{ij}(\mathbf{x}) \det[M + \Delta M_i + (F_0 + \Delta F_j) \odot F(\mathbf{x})],
\end{equation}
where $c_{ij}(\mathbf{x})$ are configuration-dependent weights; $M$ and $F_0$ denote full-rank base matrices; $\Delta M_i$ and $\Delta F_j$ represent low-rank shifts that introduce the multi-determinant character; and $F(\mathbf{x})$ is the output of the neural network backflow transformation.

Instead of using fixed mean-field orbitals, both $M$ and $F_0$ are treated as learnable parameter matrices to provide greater variational flexibility. Each of the $\Delta$ matrices is constructed as the product of two low-rank learnable matrices $U \in \mathbb{R}^{2N_\text{site} \times r}$ and $V \in \mathbb{R}^{r \times N_\text{electron}}$; that is, $\Delta M_i = U^M_i V^M_i$ and $\Delta F_j = U^F_j V^F_j$. Here the rank $r$ is chosen to be small ($r \ll N_\text{electron}$) to maintain parameter efficiency. We use the same value of $r$ for both $\Delta M_i$ and $\Delta F_j$ matrices. In practice, we also include the zero matrices $\Delta M_0 = \Delta F_0 = \mathbf{0}$, thereby retaining the original backflow-transformed determinant in the multi-determinant expansion.

The configuration-dependent weights are generated via a simple one-layer network. Specifically, average pooling is performed along the spatial dimension of the encoded variable $\mathbf{x}_{\text{encoded}}$ before the features are passed through a linear layer. Finally, a softmax activation function is applied to produce the normalized coefficients $c_{ij}(\mathbf{x})$. This normalization effectively treats the expansion as a weighted superposition of determinant components of the same state.

\subsection{Efficient determinant evaluation via the matrix determinant lemma}
\label{app:eff_det_cal}
Our multi-determinant wavefunction requires us to calculate the determinant $$\det \left[ M + \Delta M_i + \left( F_0 + \Delta F_j \right) \odot F(\mathbf{x}) \right]$$ for all $\Delta M_i$ and $\Delta F_j$. The main computational bottleneck is the matrix determinant calculation, whose number scales as the product of the number of low-rank matrices $N_{\Delta M} \times N_{\Delta F}$.

Using the matrix determinant lemma \cite{ding2007eigenvalues}, we can significantly accelerate these calculations. The matrix determinant lemma states that, for an invertible matrix $A$ of shape $n \times n$ and matrices $U$ and $V$ of shape $n \times r$, the following relation holds:
\begin{equation}
    \det(A + UV^\top) = \det(A) \det(I_r + V^\top A^{-1} U)
\end{equation}
By construction, our matrix $\Delta M_i$ is the product of two low-rank matrices $\Delta M_i = U^M_i V^M_i$. Defining $A_j = M + \left( F_0 + \Delta F_j \right) \odot F(\mathbf{x})$, with $A_j$ restricted to occupied rows, we can apply the matrix determinant lemma with $A = A_j$, $U = U^M_i$, and $V^\top = V^M_i$:
\begin{align}
    &\det \left[ M + \Delta M_i + \left( F_0 + \Delta F_j \right) \odot F(\mathbf{x}) \right] \nonumber \\
    &= \det \left[ A_j + U_i^M V_i^M \right] \\
    &= \det \left[ A_j \right] \det\left[ I_r + V_i^M A_j^{-1} U_i^M  \right].
\end{align}
Rather than calculating the full $N_{\Delta M} \times N_{\Delta F}$ determinants, we now only need to compute the $N_{\Delta F}$ determinants of the matrices $A_j$. 

In addition to $\det[A_j]$, the new formulation requires to calculate the inverse of each $A_j$. In practice, modern numerical libraries such as NumPy and JAX treat the matrix determinant and the matrix inverse as independent operations. However, both calculations fundamentally rely on an initial matrix factorization, such as the LU decomposition. After this factorization is obtained, the determinant and the required solves are cheaper than forming a full inverse: the determinant $\det[A_j]$ is simply the product of the diagonal elements of $U$ (adjusted by the pivot sign), and $A_j^{-1}U_i^M$ is found by forward and backward substitution at cost $O(N^2_{electron}r)$. Since the $O(N_\text{electron}^3)$ factorization step constitutes the primary computational bottleneck, while the subsequent products and substitutions are significantly less demanding, calling these functions separately results in redundant and costly decompositions. By explicitly performing a single factorization, we can reuse the factors for the determinant and all the required solves.

\begin{algorithm}
\linespread{1.35}\selectfont
\caption{Efficient Multi-Determinant Evaluation via Low-Rank Updates}
\label{alg:multi_det}
Inputs: 
$M, F_0, U^M_i, V^M_i, U^F_j, V^F_j, F(\mathbf{x})$ \;

\For{j $\gets$ 1 to $N_{\Delta F}$}{
$\Delta F_j \leftarrow U_j^F V_j^F$ \;
$A_j \gets M + \left( F_0 + \Delta F_j \right) \odot F(\mathbf{x})$ \;
$[L_j, U_j, P_j] \leftarrow \text{LU\_Decomposition}(A_j)$ \;
$\det(A_j) \leftarrow \text{det\_from\_LU}(L_j, U_j, P_j)$ \;
    \For{ i $\gets$ 1 to $N_{\Delta M}$}{
        $X_{ij} \gets \text{LU\_Solve}(L_j, U_j, P_j, U^M_i)$ \;
        $B_{ij} \gets I_r + V^M_i X_{ij}$ \;
        $\psi_{ij} = \det(A_j) \det(B_{ij})$ \;
    } 
}
\Return{Vector $\psi_{ij}$}
\end{algorithm}

The remaining computation is the matrix multiplication and the determinant calculation of $\det\left[ I_r + V_i^M A_j^{-1} U_i^M  \right]$. The former has a complexity of $\mathcal{O}(N_\text{electron}^2)$, and the latter scales as $\mathcal{O}(r^3)$. Given that $r \ll N_\text{electron}$, these computations are significantly cheaper than the original $\mathcal{O}(N_\text{electron}^3)$ determinant calculation. Therefore, we can reduce the overall computational cost from $\mathcal{O}(N_{\Delta M}N_{\Delta F}N_\text{electron}^3)$ to $\mathcal{O}(N_{\Delta F}N_\text{electron}^3) + \mathcal{O}(N_{\Delta M}N_{\Delta F}N_\text{electron}^2r)$. Algorithm $\ref{alg:multi_det}$ describes the efficient multi-determinant calculation that utilizes the low-rank updates.

\begin{figure}[ht]
    \includegraphics[scale=0.39]{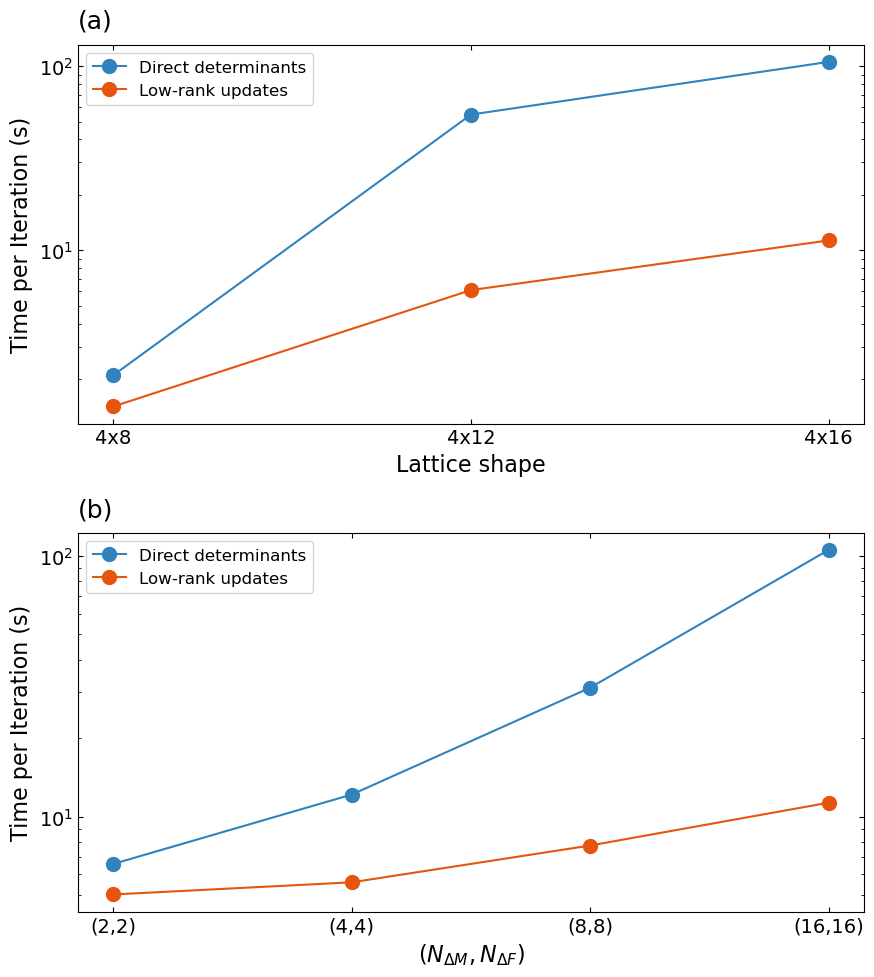}
    \centering
    \caption{Scaling of computational cost for direct determinant evaluation versus the low-rank lemma application with $r=2$. (a) Training time per iteration as a function of the lattice sizes for a fixed nonzero shift counts $(16,16)$. (b) Training time per iteration relative to the nonzero shift counts $(N_{\Delta M}, N_{\Delta F})$ for a fixed $4 \times 16$ lattice. Both plots utilize a logarithmic scale for the vertical axis.
    }
    \label{fig:time_comparison}
\end{figure}

We tested the speed improvement of the low-rank updates at $1/8$ doping with $U/t = 8$. Fig.~\ref{fig:time_comparison} compares the training time per iteration between direct determinant evaluation and the low-rank update method. We ran the experiments on a single NVIDIA H200 GPU. Panel (a) shows that as the lattice size increases, the training time for the direct determinant method rises from roughly 2 seconds to over 100 seconds. Over the same range, the time for the low-rank updates increases from approximately 1.5 seconds to 11 seconds. At the largest system size, the direct method is approximately 10 times slower than the low-rank approach. In panel (b), increasing the nonzero shift counts from $(2,2)$ to $(16,16)$ results in the direct determinant time rising from approximately 6 seconds to 105 seconds, while the low-rank method reaches only approximately 11 seconds for shift counts $(16,16)$.

\section{Energy benchmarks and convergence}
We test our neural network wavefunction ansatz on the 2D Hubbard model for both the half-filling and 1/8 doping cases, with periodic boundary conditions applied on both spatial directions. For both cases, we test on the strong correlation regime where $U/t = 8$. After training convergence, the variational energy and its associated error bar are calculated by generating 10 independent Monte Carlo runs, each consisting of 16,384 samples. These runs measure sampling uncertainty at fixed parameters, not variation between training runs.

\begin{table*}[ht]
    \centering
    \caption{Comparison of different network architectures. The last column reports total energies.}
    \label{tab:network_models_table}
    \begin{tabular*}{\textwidth}{@{\extracolsep{\fill}}llcccc@{}}
        \toprule
        \textbf{Model} & \textbf{Backflow} & \textbf{Coefficients} & \textbf{Embedding Dim. (d)} & \textbf{\# Params} & \textbf{Energy ($E$)} \\ 
        \midrule
        Net 1 & Simple & Independent & 32 & 44,861 & $-24.394$ \\
        Net 2 & Simple & Dependent   & 32 & 54,109 & $-24.419$ \\
        Net 3 & Deep   & Dependent   & 32 & 61,317 & $-24.464$ \\
        Net 4 & Deep   & Dependent   & 40 & 81,693 & \textbf{$-24.474$} \\ 
        \bottomrule
    \end{tabular*}
\end{table*}

\subsection{Ground state energy}
\begin{figure}[ht]
    \includegraphics[scale=0.3]{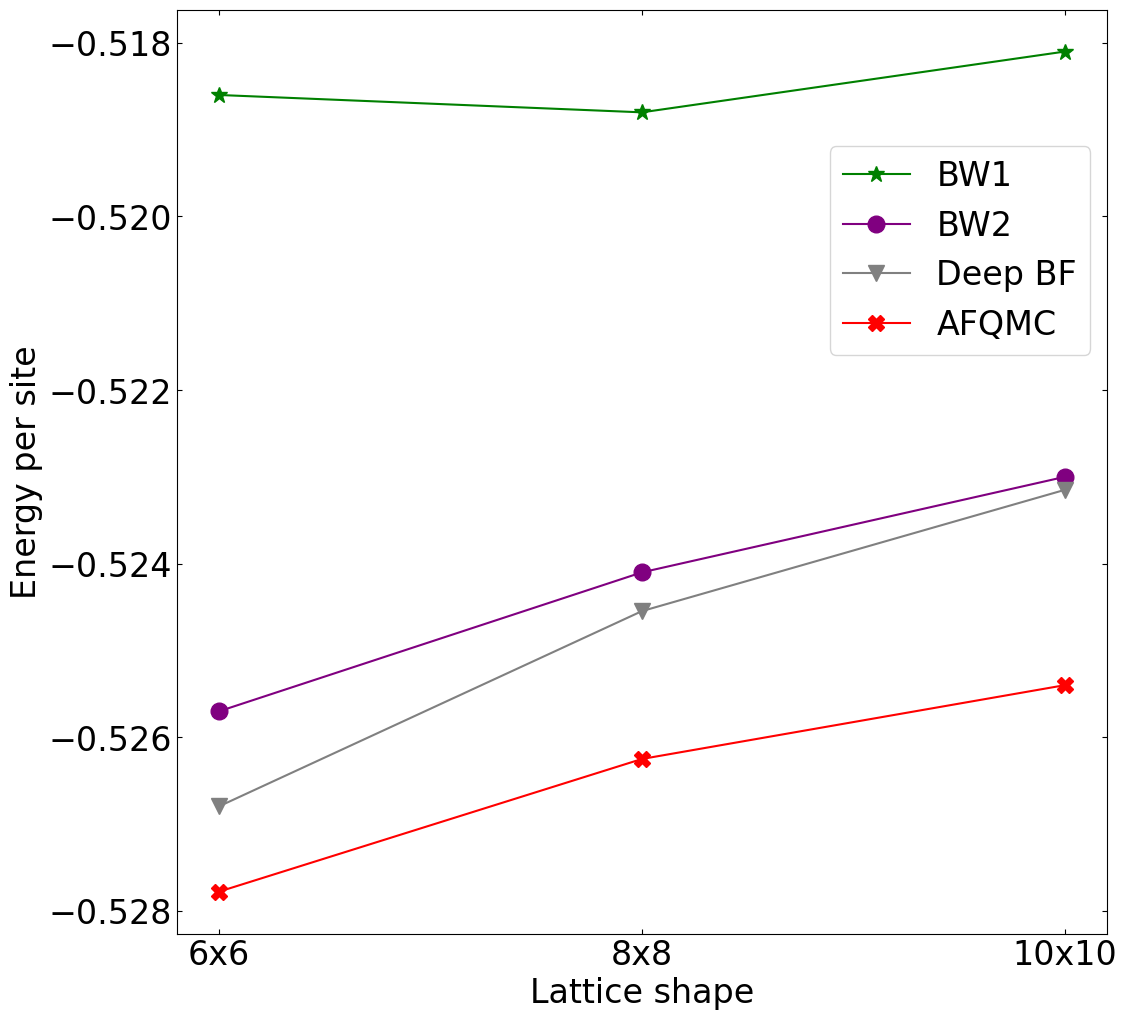}
    \centering
    \caption{Variational energy per site for the 2D Hubbard model at half-filling with respect to different lattice sizes. The plot compares the performance of our deep backflow (Deep BF) to various neural network backflow architectures, including  spin-block-separated (BW1) and spin-mixing (BW2) tensor backflow, both with one Lanczos step \cite{zhou2024solving}, and the auxiliary-field quantum Monte Carlo (AFQMC) benchmarks \cite{qin2016benchmark} for different $L \times L$ lattice sizes. The Deep BF error bars are smaller than the markers.
    }
    \label{fig:energy_half_filled}
\end{figure}

To evaluate the accuracy of the proposed architecture, we first consider the 2D Hubbard model at half-filling. Fig.~\ref{fig:energy_half_filled} illustrates the variational energy per site across different square lattices. 
We adopt the energy results from AFQMC \cite{qin2016benchmark} as the numerical benchmarks for comparison. On the $6 \times 6$ lattice, the Deep BF model achieves a variational energy of $-0.5268$ per site, lying below the BW1 and BW2 point estimates. For the intermediate $8 \times 8$ geometry, Deep BF yields an energy of $-0.5245$ per site, again lying below the BW1 and BW2 point estimates. On the largest $10 \times 10$ lattice, the Deep BF result reaches $-0.5231$ per site, lying below BW1 and very close to BW2.

\begin{figure}[ht]
    \includegraphics[scale=0.3]{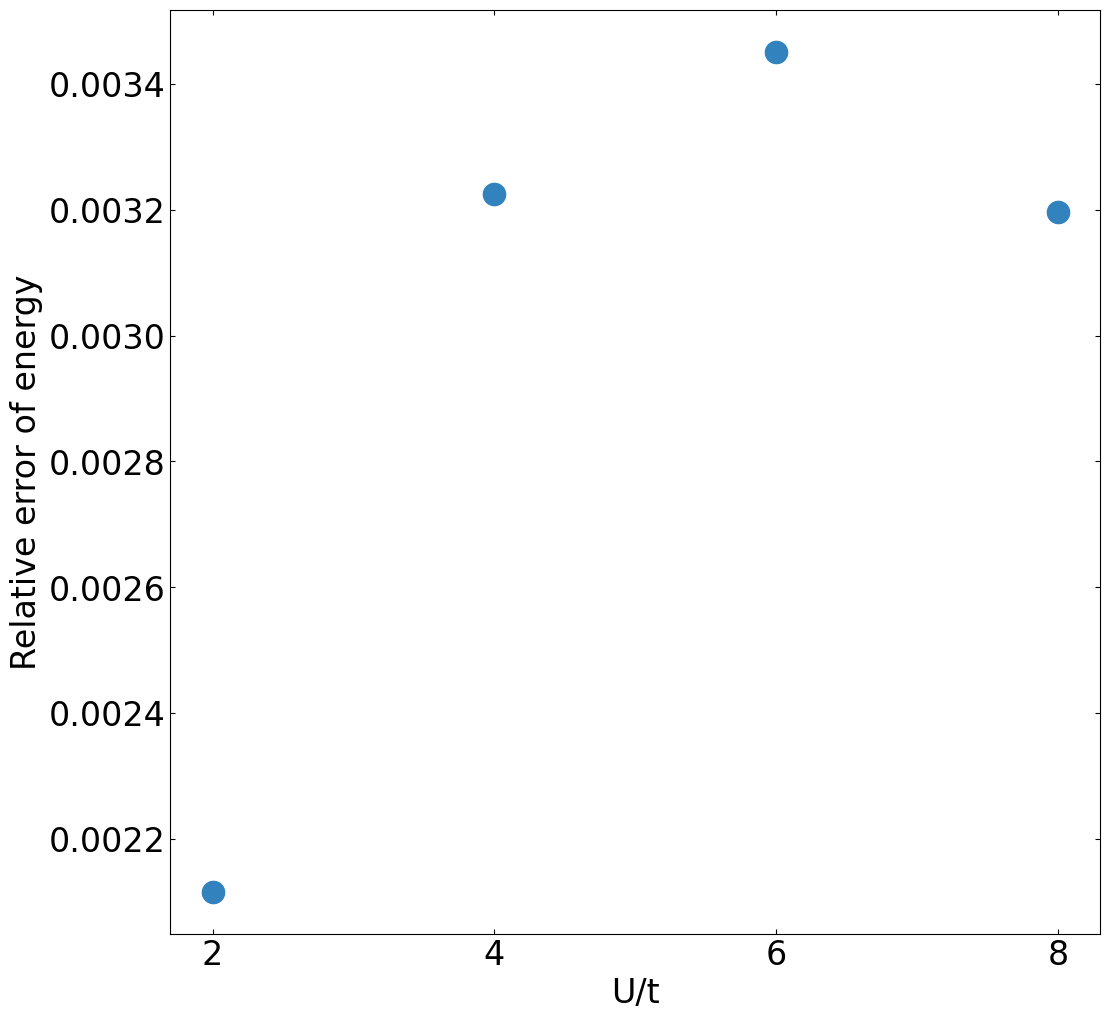}
    \centering
    \caption{Relative error of the variational energy for the $8 \times 8$ Hubbard model at half-filling with respect to different $U/t$ values. The relative error is calculated with respect to the AFQMC benchmarks \cite{qin2016benchmark}.
    }
    \label{fig:energy_half_fill_8x8_different_U}
\end{figure}
To assess the performance of our network architecture across different correlation regimes, we analyze the relative error of the variational energy as a function of the on-site interaction strength $U$ at half-filling on the $8 \times 8$ lattice. As shown in Fig.~\ref{fig:energy_half_fill_8x8_different_U}, the relative error is computed directly against the AFQMC ground-state energy benchmark. In the weakly correlated regime at $U/t=2$, the Deep BF method achieves its highest precision, yielding a remarkably low relative error of approximately $0.0021$. As the interaction strength increases into the intermediate coupling regime, the relative error rises to approximately $0.0032$ at $U/t=4$ and peaks at roughly $0.0034$ near $U/t=6$. Interestingly, as the system enters the strongly correlated regime at $U/t=8$, the trend reverses, and the relative error decreases slightly to approximately $0.0032$.

\begin{figure}[ht]
    \includegraphics[scale=0.3]{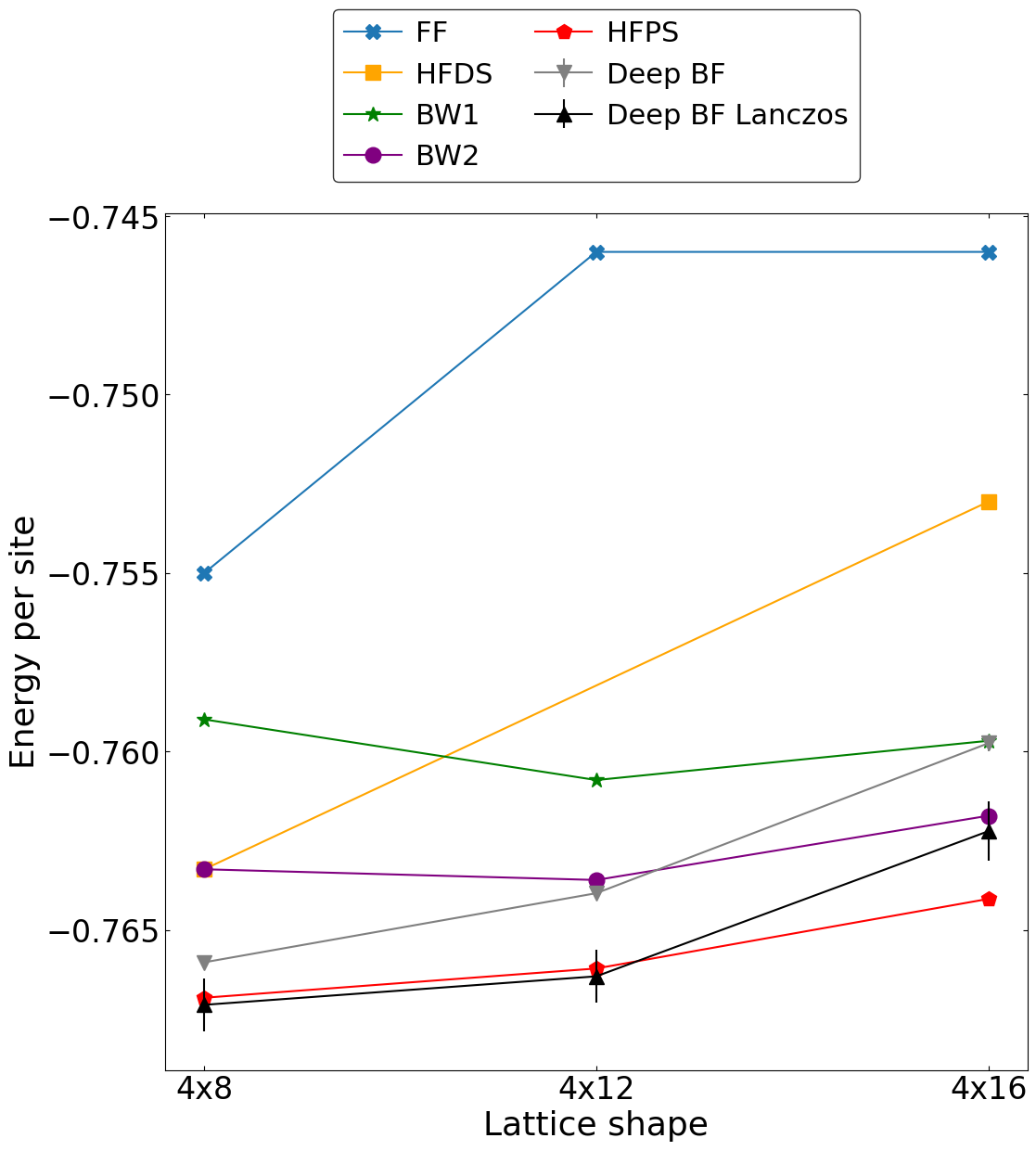}
    \centering
    \caption{Variational energy per site for the 2D Hubbard model at 1/8 doping. The plot compares our proposed deep backflow network with feed-forward backflow (FF) \cite{luo2019backflow}, hidden fermion determinant state (HFDS) \cite{robledo2022fermionic}, tensor backflow (BW1 and BW2) \cite{zhou2024solving}, and hidden fermion Pfaffian state (HFPS) \cite{chen2025neural} for different $4 \times L$ lattice sizes. The gray triangles represent our deep backflow result, and the black triangles represent the deep backflow result with a Lanczos step. The deep backflow error bars are smaller than the markers.
    }
    \label{fig:energy_doped}
\end{figure}
We further extend our analysis to the $1/8$ doped regime. To assess the quality of the optimized variational wave function $|\Psi\rangle$, we apply a one-step Lanczos correction after training. This procedure enlarges the variational space to the Krylov subspace spanned by $|\Psi\rangle$ and $\hat{H}|\Psi\rangle$, yielding the corrected state
\begin{equation}
|\Psi_{\mathrm{L}}\rangle = (1+\alpha \hat{H})|\Psi\rangle ,
\end{equation}
where $\alpha$ is a variational parameter. The optimal $\alpha$ is obtained by solving the associated $2\times2$ generalized eigenvalue problem for the projected Hamiltonian and overlap matrices and selecting the lowest eigenvalue solution. All required matrix elements are estimated from Monte Carlo samples drawn from the optimized reference state $|\Psi\rangle$. The energy variance after the Lanczos step requires higher-order moments up to $\langle \hat{H}^4 \rangle$, which are computed using the same stochastic estimators.

Fig.~\ref{fig:energy_doped} presents the training results for the doped case, where a network embedding dimension of 64 was used. On the $4 \times 8$ lattice, the Deep BF result reaches an energy of $-0.7659$ per site, which is higher than only HFPS and Deep BF Lanczos,  the latter of which yields the lowest energy point estimate on this geometry at $-0.7671$ per site. For the $4 \times 12$ system, the Deep BF energy is $-0.7640$ per site, placing it below FF, BW1 and BW2, but above HFPS and Deep BF Lanczos. Deep BF Lanczos achieves an energy of $-0.7662$ per site, sitting slightly below HFPS. On the largest $4 \times 16$ lattice, the Deep BF model achieves an energy of $-0.7598$ per site, outperforming FF and HFDS. Deep BF Lanczos finishes at $-0.7622$ per site, which lies slightly below BW2 but above HFPS.

To evaluate the impact of architectural complexity on variational accuracy, we designed four neural network models with varying degrees of flexibility in their backflow transformations and determinant coefficients (Table \ref{tab:network_models_table}). We tested these models on a $4 \times 8$ lattice at $1/8$ doping with $U/t = 8$. The models range from Net 1, a baseline architecture with simple backflow and independent coefficients, to Net 4, a high-capacity model utilizing deep backflow transformations and a larger embedding dimension. This hierarchy allows us to perform a variance extrapolation to estimate the exact ground-state energy. Furthermore, these configurations serve as the basis for the architectural analysis and interpretability studies conducted in the following sections.

\begin{figure}[ht]
    \includegraphics[scale=0.37]{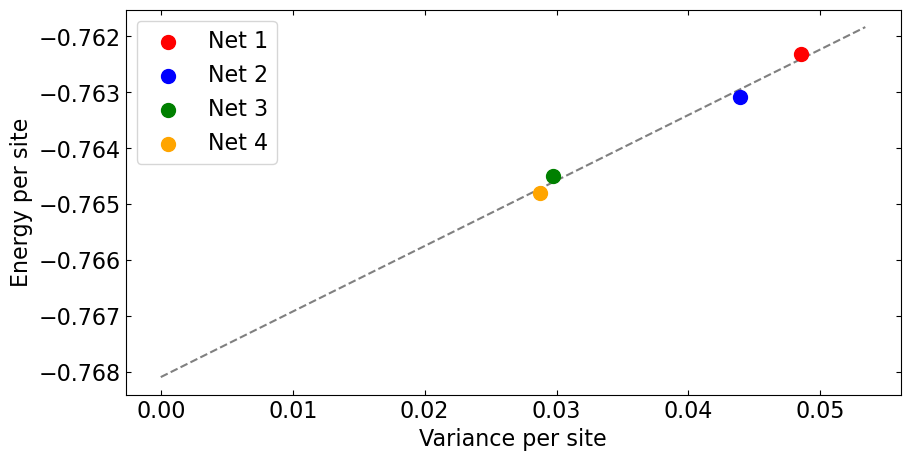}
    \centering
    \caption{Variance extrapolation of the ground-state energy for the 4x8 lattice. The energy per site for the four network architectures (Net 1–4) is plotted against the local energy variance per site. The dashed gray line represents a linear fit to the data. The $y$-intercept indicates the extrapolated ground-state energy in the limit of zero variance, yielding $E_{\text{extrap}} = -0.7681$.
    }
    \label{fig:variance_extrapolate}
\end{figure}

As shown in Fig.~\ref{fig:variance_extrapolate}, our models exhibit a consistent linear trend. The systematic reduction in variance, achieved by transitioning from simple backflow (Net 1) to a deeper, higher-dimensional transformer architecture (Net 4), directly corresponds to a lowering of the variational energy. By performing a linear least-squares fit to these data points and extrapolating to the zero-variance limit, we arrive at an estimated ground-state energy of $-0.7681$ per site.

\subsection{Hyperparameter sensitivity}
\begin{figure}[ht]
    \includegraphics[scale=0.48]{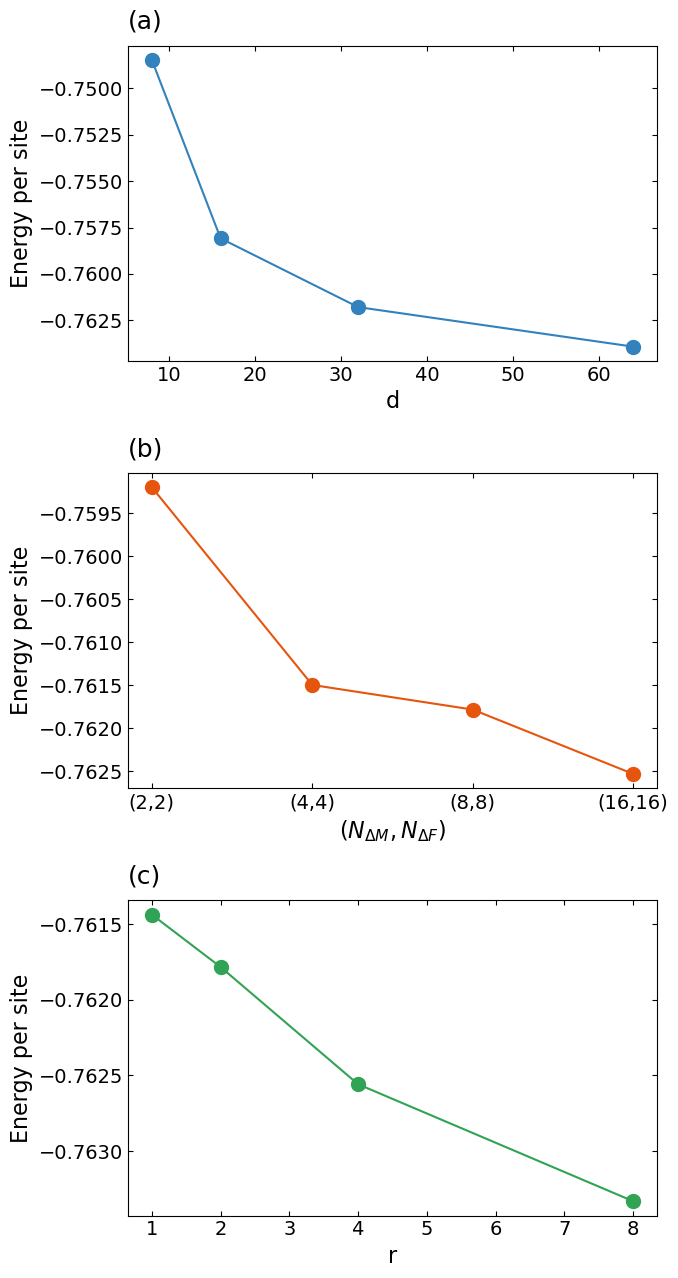}
    \centering
    \caption{Energy per site for the $4 \times 8$ lattice at $1/8$ doping as a function of key model hyperparameters. (a) Network embedding dimension. (b) Number of additional low-rank matrices $(N_{\Delta M}, N_{\Delta F})$. (c) Rank $r$ of the additional $\Delta$ matrices.
    }
    \label{fig:convergence_plots}
\end{figure}

To assess sensitivity of our variational results, we examine the dependence of the ground-state energy with respect to the key hyperparameters of the neural backflow ansatz. As shown in Fig.~\ref{fig:convergence_plots}, we evaluate the energy per site for the $4 \times 8$ lattice at $1/8$ hole doping with $U/t=8$. We define our reference setup for these experiments as $d=32$, $(N_{\Delta M}, N_{\Delta F})=(16,16)$, and $r=2$.

\begin{figure*}[ht]
    \centering
    \includegraphics[scale=0.48]{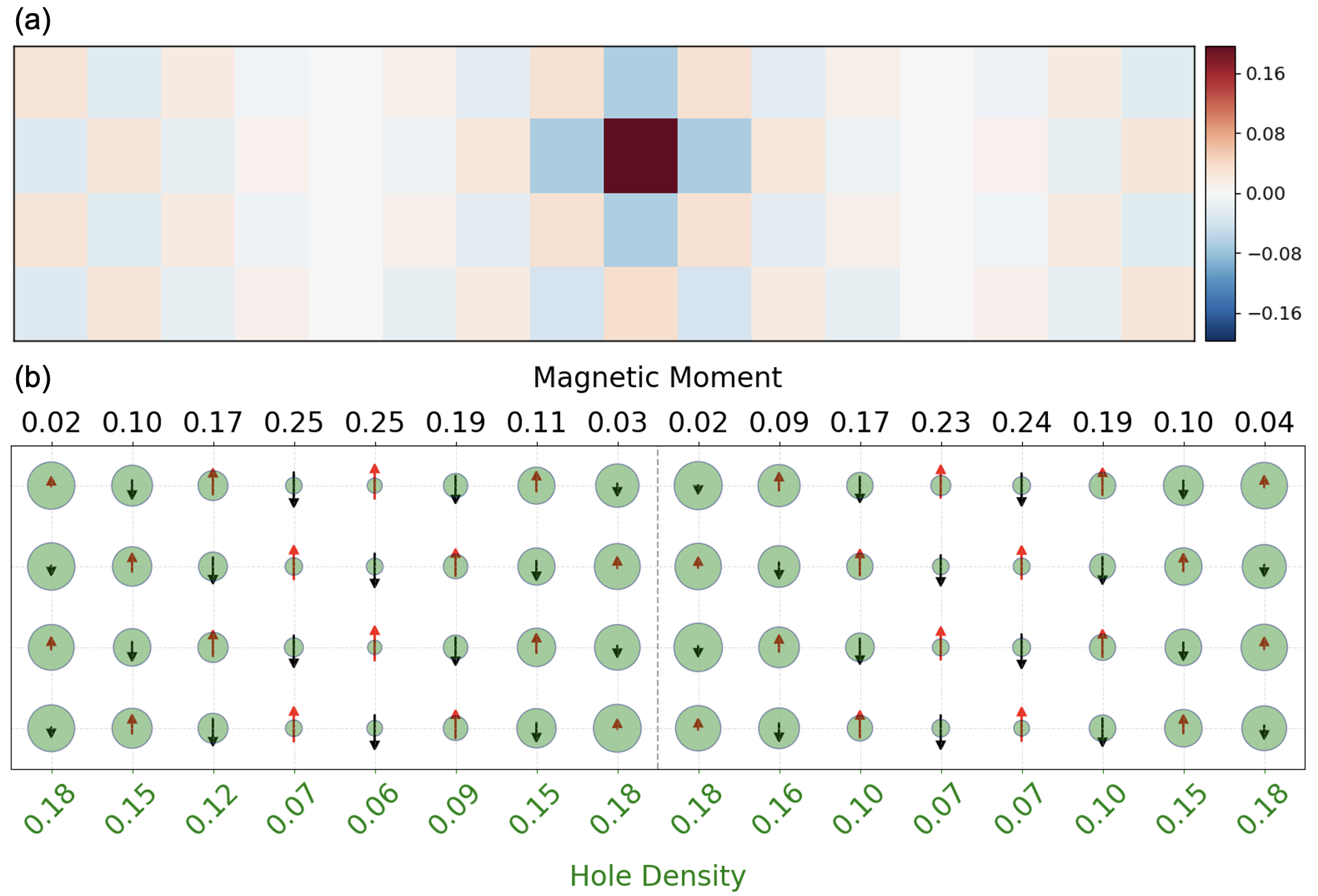}
    \caption{Real-space behaviors of the doped Hubbard model. (a) Real-space spin-spin correlation function $C_s(\mathbf{r})$ with the origin of the displayed displacement at the center. (b) Snapshot of local observables; the radius of green circles represents the local hole density $h_i$ and the arrows represent the local magnetic moment $m_i$. The lower values are column-averaged hole densities, while the upper values describe moment magnitudes.}
    \label{fig:real_space_observables}
\end{figure*}
Specifically, panel (a) illustrates the dependence on the embedding dimension for $d=8, 16, 32, 64$. The energy per site exhibits a sharp decrease between $d=8$ and $d=16$, followed by smaller energy gains as $d$ approaches 64. In panel (b), we observe the impact of the determinant expansion $(N_{\Delta M}, N_{\Delta F})$ at values of $(2,2), (4,4), (8,8),$ and $(16,16)$. The plot shows a consistent monotonic reduction in energy across the entire range, with the most significant improvement occurring between the $(2,2)$ and $(4,4)$ configurations. Finally, panel (c) shows the sensitivity to the low-rank cutoff for $r=1, 2, 4, 8$. The energy per site displays a steady, nearly linear decrease as the rank is increased.

\section{Physics of the Doped State}
We investigate the optimized variational-state properties of the $1/8$-doped $4 \times 16$ Hubbard model at $U/t=8$ using a multi-scale analysis. We begin with local real-space observables of spin and charge, followed by an analysis of collective behavior and pairing tendencies via structure factors.

\subsection{Real-Space Stripe Pattern}
We first examine real-space signatures of stripe order through local observables on the $4 \times 16$ lattice. Specifically, we compute the spin--spin correlation
\begin{equation}
C_s(\mathbf{r}) = \frac{1}{N} \sum_{i} \langle \hat{S}^z_i \hat{S}^z_{i+\mathbf{r}} \rangle,
\end{equation}
the local hole density
\begin{equation}
h_i = 1 - \langle \hat{n}_i \rangle,
\end{equation}
and the local magnetic moment
\begin{equation}
m_i = \frac{1}{2} \left( \langle \hat{n}_{i\uparrow} \rangle - \langle \hat{n}_{i\downarrow} \rangle \right).
\end{equation}

Fig.~\ref{fig:real_space_observables}(a) displays $C_s(\mathbf{r})$, revealing a robust but spatially modulated antiferromagnetic (AFM) background. Notably, we observe two distinct domain walls characterized by a sign flip in the AFM order. These magnetic boundaries align precisely with regions of high hole concentration, as further elucidated in panel (b). The visualization reveals that the high-density regions ($h_i \approx 0.18$) at the lattice boundaries and the central column align perfectly with the magnetic domain walls. In these regions, the local magnetic moments are suppressed to a minimum of $0.03$. Conversely, in the lower-density regions ($h \approx 0.06$), the magnetic moments recover to values as high as $0.25$, maintaining a standard checkerboard pattern. These observations indicate the synchronization of a spin-density wave (SDW) and a charge-density wave (CDW).

\subsection{Momentum-Space Correlations and Pairing}
The collective behavior of the system is further characterized by the spin structure factor 
\begin{equation}S(\mathbf{q}) = \frac{1}{N_\text{site}} \sum_{j,k} e^{i\mathbf{q}\cdot(\mathbf{r}_j - \mathbf{r}_k)} \langle \hat{S}^z_j \hat{S}^z_k \rangle
\end{equation}
and the charge structure factor 
\begin{equation}
    \quad N(\mathbf{q}) = \frac{1}{N_\text{site}} \sum_{j,k} e^{i\mathbf{q}\cdot(\mathbf{r}_j - \mathbf{r}_k)} (\langle \hat{n}_j \hat{n}_k \rangle - \langle \hat{n}_j \rangle \langle \hat{n}_k \rangle).
\end{equation}

\begin{figure}[ht]
\centering
\includegraphics[scale=0.48]{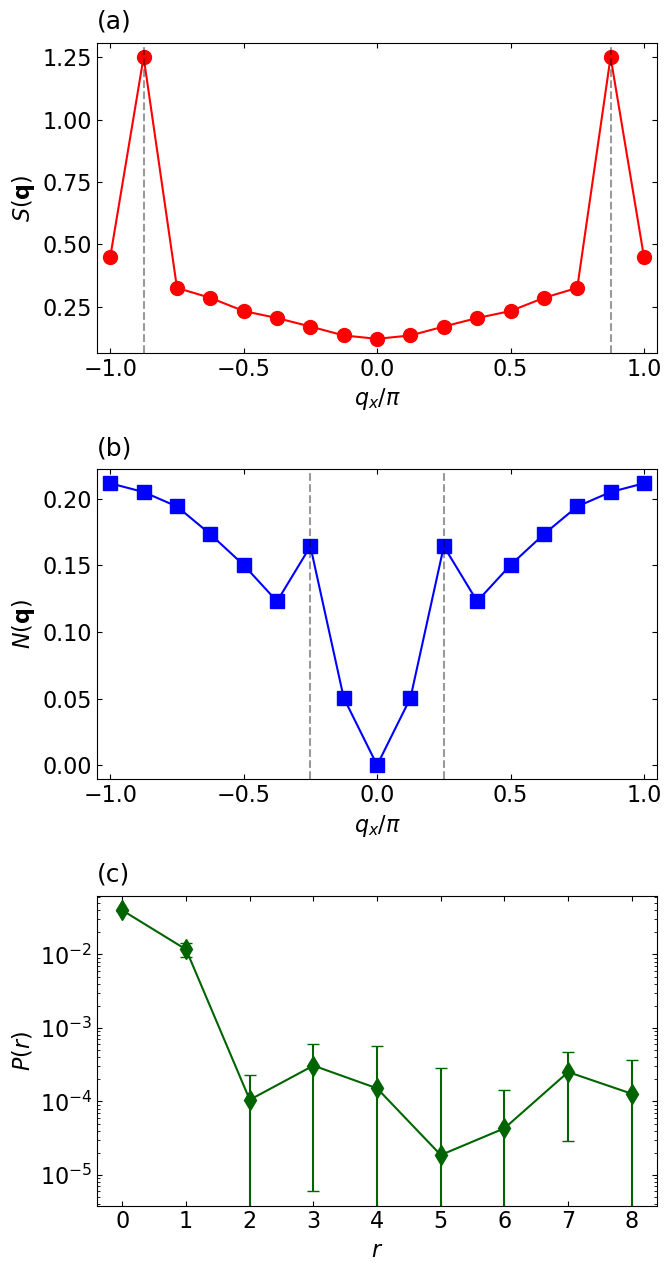}
\caption{Computed observables for the $4 \times 16$ Hubbard model at $1/8$ doping. (a) $S(\mathbf{q})$ showing incommensurate SDW peaks at $q_x = \pm 7\pi/8$. (b) $N(\mathbf{q})$ displaying CDW peaks at $q_x = \pm \pi/4$. (c) Pair-pair correlation $P(r)$ along the long axis on a semi-log scale. Error bars denote the statistical uncertainty from Monte Carlo sampling; their lower portions are truncated due to the logarithmic scale. }
\label{fig:correlations}
\end{figure}

As presented in Fig.~\ref{fig:correlations}(a), the spin structure factor $S(\mathbf{q})$ exhibits prominent symmetric peaks at $q_x = \pm 7\pi/8$. This incommensurate shift from the AFM point ($\pi$) is consistent with the magnetic periodicity observed in real space. In panel (b), $N(\mathbf{q})$ shows distinct CDW peaks located at $q_x = \pm \pi/4$.

We examine long-range pairing tendencies via the pair--pair correlation function defined as
\begin{equation}
P(\mathbf{r}) = \frac{1}{N_{\text{site}}} \sum_{i} 
\left\langle \hat{\Delta}_i^\dagger \, \hat{\Delta}_{i+\mathbf{r}} \right\rangle
\end{equation}
where $\hat{\Delta}_i^\dagger$ is the creation operator for a singlet Cooper pair with $d_{x^2-y^2}$ symmetry anchored at site $i$:
\begin{equation}
\hat{\Delta}_i^\dagger = \frac{1}{\sqrt{2}} \sum_{\boldsymbol{\delta}} f(\boldsymbol{\delta}) 
\left(
\hat{c}_{i\uparrow}^\dagger \hat{c}_{i+\boldsymbol{\delta},\downarrow}^\dagger
-
\hat{c}_{i\downarrow}^\dagger \hat{c}_{i+\boldsymbol{\delta},\uparrow}^\dagger
\right),
\end{equation}
where $\boldsymbol{\delta} \in \{ \pm\hat{x}, \pm\hat{y} \}$ denotes the nearest-neighbor vectors, and the form factor $f(\boldsymbol{\delta})$ is defined as $+1$ for bonds along the $x$-direction and $-1$ for bonds along the $y$-direction. Panel (c) plots $P(r)$ on a semi-logarithmic scale for distances $r \in [0, 8]$. The pair correlations beyond the first shell are zero to within statistical uncertainty. The correlation magnitude is highest at $r=0$ and undergoes a sharp decrease of approximately two orders of magnitude by $r=2$. For $r > 2$, $P(r)$ exhibits oscillatory behavior with local maxima at $r=3$ and $r=8$, fluctuating within the range of $10^{-3}$ to $10^{-4}$.

\section{Architectural analysis and interpretability}
\subsection{Static kernel heatmaps}
To gain insight into how the neural network captures the underlying lattice structure, we examine the learned weights of the static kernels $\alpha$ across different layers of the encoder. Unlike the dynamic components of the network that respond to specific electron configurations, these static kernels represent an input-independent mapping that enforces global spatial correlations. In a physical context, these kernels act as real-space filters that mix features
across lattice patches. By visualizing these weights as heatmaps, we can directly observe the geometric prior the model develops during the variational optimization process. 

\begin{figure}[ht]
    \includegraphics[scale=0.225]{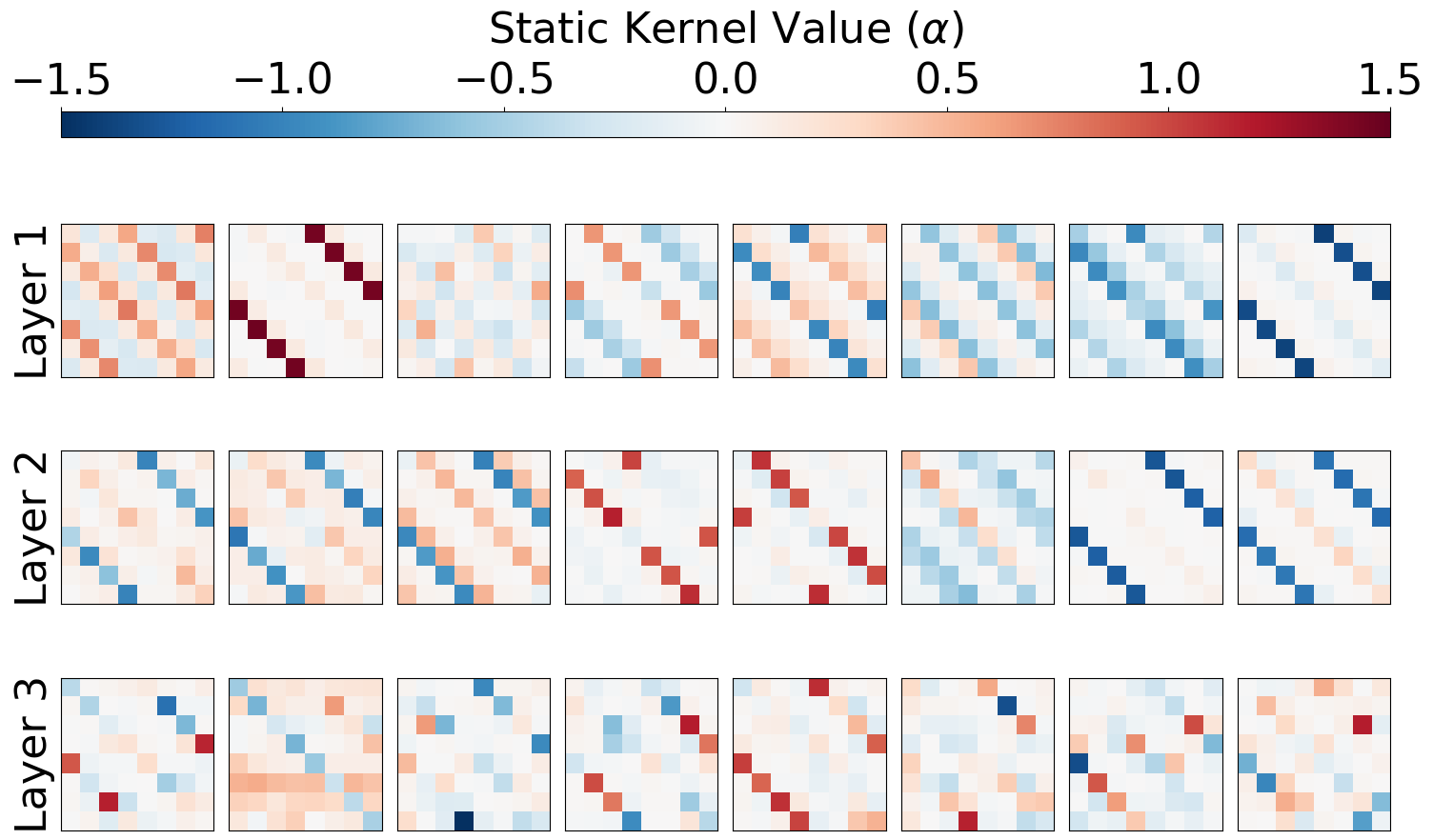}
    \centering
    \caption{Learned static kernels $\alpha$ across 3 encoder layers, each with 8 independent kernels. The heatmaps show how the model learns to map spatial relationships on the lattice, with red and blue representing positive and negative patch-mixing weights. 
    }
    \label{fig:kernel_heatmap}
\end{figure}

Fig.~\ref{fig:kernel_heatmap} plots the static kernel heatmap for different numbers of layers and independent kernels trained on the $4\times8$ lattice at 1/8 doping with PBCs applied on both dimensions. The first encoder layer serves as a geometric preprocessor, mapping the input sequence onto a 2D lattice topology. Despite the linear indexing of patches, the network identifies wrap-around connectivity by placing significant weight on the off-diagonal corners, providing evidence for learned periodic boundary conditions. These learned interactions remain highly localized, prioritizing the nearest-neighbor connections necessary to optimize the system's kinetic energy. 

In the second layer, the kernels show more diagonal structure, marked by the emergence of strong diagonal elements that may reflect self-interactions within each patch. This diagonal dominance suggests that the encoder may be placing greater emphasis on local renormalization, potentially reflecting the need to account for the high energy costs associated with electron-electron interactions. By the final layer, the kernels transition into a more complex, stochastic distribution, with the strict symmetries observed in earlier stages no longer apparent. The absence of a simple real-space pattern may be consistent with the representation of higher-order many-body correlations.

\subsection{Determinant-basis orthogonality}
The accuracy of a neural-network backflow wavefunction depends not only on the expressive power of the network layers but also on the quality and flexibility of the multi-determinant expansion it generates. To elucidate the physical structure learned by the various architectures, we perform a systematic analysis of the determinant basis across four distinct network setups, as summarized in Table~\ref{tab:network_models_table}. These analyses are conducted on a $4\times8$ lattice at $1/8$ doping with PBCs applied in both dimensions. To construct the multi-determinant wavefunction in Eq.~(\ref{eq:multi_dets}), we utilize an expansion set of 17 $\Delta M_i$ and 17 $\Delta F_j$ matrices, each consisting of 16 distinct low-rank factorized matrix shifts and a special zero matrix. For each network model, we generate a design matrix $X \in \mathbb{R}^{N_\text{sample} \times N_\text{det}}$, consisting of $N_\text{sample} = 4,096$ sampled test configurations as rows and $N_\text{det} = 289$ determinants as columns.

We begin with Net 1 as a baseline, featuring a simple backflow network consisting of only two linear layers and configuration-independent coefficients for the linear combination of the determinants. In Net 2, we introduce configuration-dependent coefficients to increase the flexibility of the multi-determinant expansion. Net 3 marks a significant transition to a deep backflow architecture—as illustrated in Fig.~\ref{fig:network_architectures}—which enables the capture of more complex, non-local electronic correlations. Finally, Net 4 extends the deep backflow by increasing the embedding dimension from $d=32$ to $d=40$, thereby increasing the overall expressive power of the network. As shown in the last two columns of Table~\ref{tab:network_models_table}, each architectural refinement leads to an increase in the total number of parameters while simultaneously providing a systematic lowering of the variational energy, with Net 4 yielding the most accurate ground-state approximation. 

\subsection{Gram matrices and singular value spectrum}
\begin{figure}
    \includegraphics[scale=0.281]{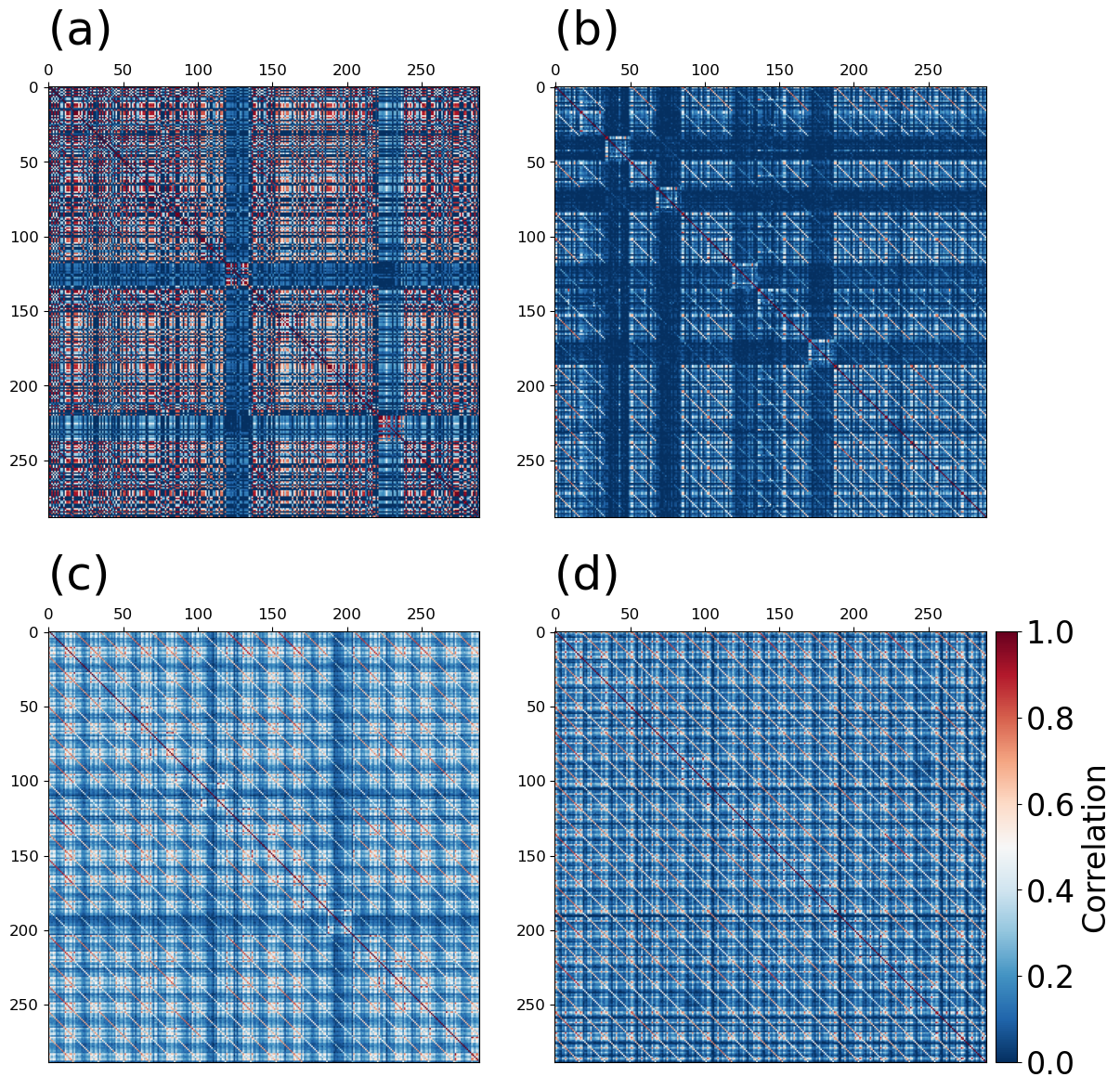}
    \centering
    \caption{Gram matrices of the network models. The heatmaps display the correlation (cosine similarity) between the 289 determinants for architectures (a) Net 1, (b) Net 2, (c) Net 3, and (d) Net 4.
    }
    \label{fig:gram_matrices}
\end{figure}

To characterize the pairwise similarity and redundancy of the learned determinant basis, we compute the Gram matrix $G \in \mathbb{R}^{N_\text{det} \times N_\text{det}}$ for each of the four network architectures. Across the sampled configurations in our design matrix, we normalize the column vectors (determinants) to unit length, such that the elements of the Gram matrix, $G_{ij} = \langle \det_i | \det_j \rangle$, represent the cosine similarity between the $i$-th and $j$-th determinant basis states.

Fig~\ref{fig:gram_matrices} displays the Gram matrices for Net 1 through 4. A notable feature present in all four Gram matrices is the recurring cross-hatch pattern. This structure is a direct consequence of the multi-determinant construction in Eq.~(\ref{eq:multi_dets}), where the basis is formed by the outer product of $17$ $\Delta M_i$ and $17$ $\Delta F_j$ shifts. The determinant basis is therefore naturally grouped into blocks of 17 that exhibit inherent structural similarities within the sampled representation.

As the network becomes deeper and more expressive, from Net 1 to Net 4, we observe an overall improvement in the energy results, together with a decrease in the density of the off-diagonal elements in the Gram matrix. This indicates that the determinant basis states exhibit less pairwise similarity in the later architectures. While this does not by itself establish linear independence or orthogonality, it may suggest that the more expressive networks learn to organize the multi-determinant basis such that the sampled determinants explore more distinct directions in the basis space. This may allow the network to make more effective use of the available multi-determinant degrees of freedom.

\begin{figure}[ht]
    \includegraphics[scale=0.281]{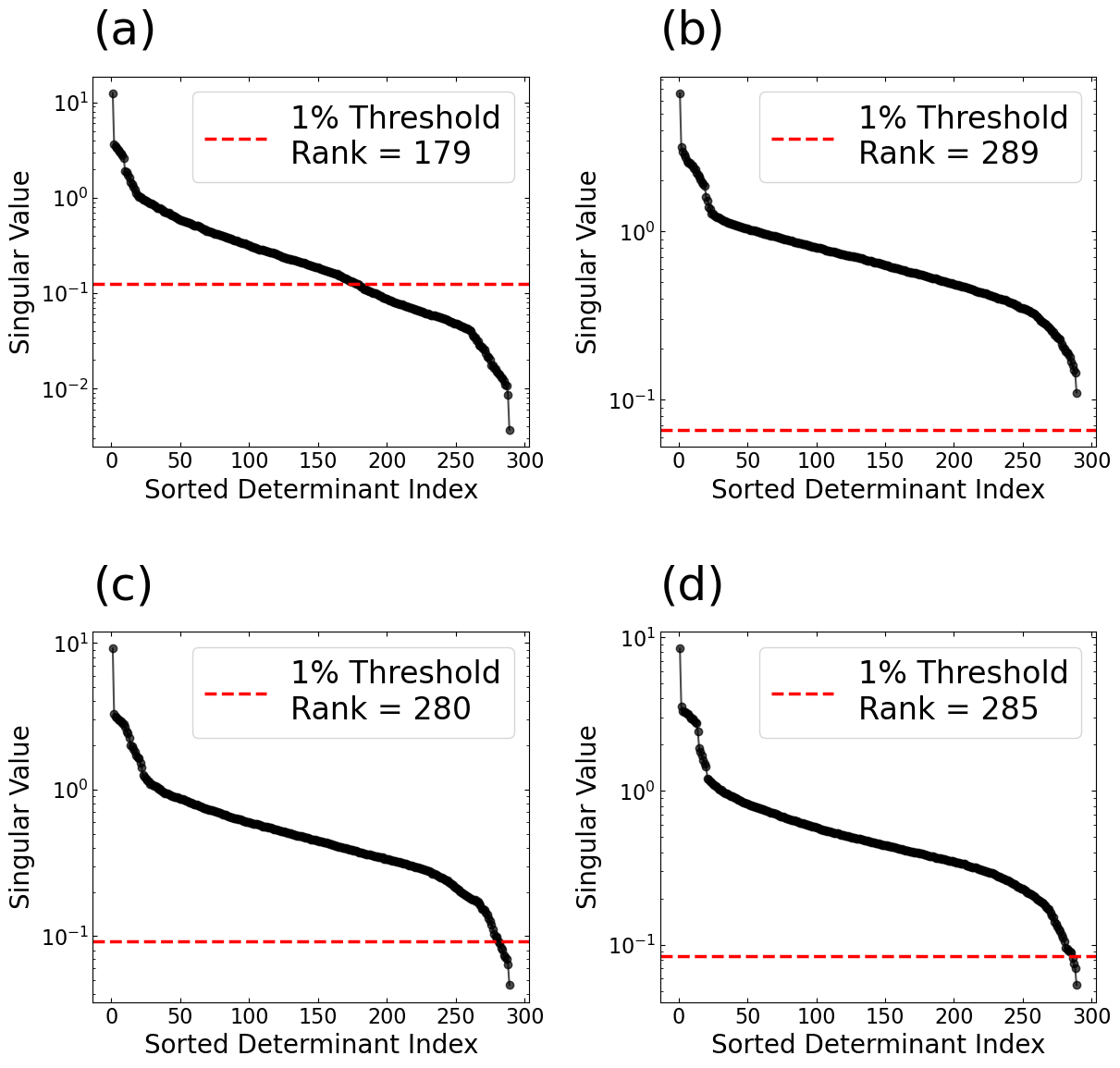}
    \centering
    \caption{Singular value spectrum of the network models in log scale. The plots show the singular values of the determinant design matrix $X$ for architectures (a) Net 1, (b) Net 2, (c) Net 3, and (d) Net 4. The red dashed line denotes a 1$\%$ threshold relative to the largest singular value, used to define the effective rank of the basis.
    }
    \label{fig:sv_plots}
\end{figure}

To further characterize the redundancy observed in the Gram matrices, we compute the singular value spectrum of the determinant basis for each network. We define the effective rank as the number of singular values exceeding $1\%$ of the maximum singular value, providing an effective measure of the dimensionality of the 289-determinant expansion.

As shown in Fig.~\ref{fig:sv_plots}, Net 1 exhibits the strongest redundancy, with an effective rank of only $179$. The introduction of configuration-dependent coefficients in Net 2 increases the effective rank to the full basis size of $289$, although the resulting energy remains higher than that of the deeper architectures. The transition to deep backflow in Net 3 is accompanied by a slight reduction in effective rank to $280$, which may suggest some consolidation of the represented features. By increasing the embedding dimension in Net 4, the effective rank increases again to $285$, while the network achieves the lowest variational energy. Across all four networks, the singular value spectra also exhibit a similar three-stage decay: an initial rapid drop, followed by a relatively stable intermediate region, and finally a more gradual decay toward the basis limit. This structure suggests that the determinant basis contains a small number of dominant modes together with a broader set of less dominant modes that contribute to the represented many-body state.

\subsection{Principal component analysis of the determinant basis}
To further examine how the determinant basis represents the sampled configurations, we employ principal component analysis (PCA) to visualize the high-dimensional configuration space. We project the $4,096$ sampled test configurations from the original $289$-dimensional determinant space onto the first two principal components (PC1 and PC2), which correspond to the directions of largest variance in the data. This projection provides a low-dimensional view of the variation among the sampled configurations and allows us to compare how different network architectures organize the configuration space.

\begin{figure}
    \includegraphics[scale=0.29]{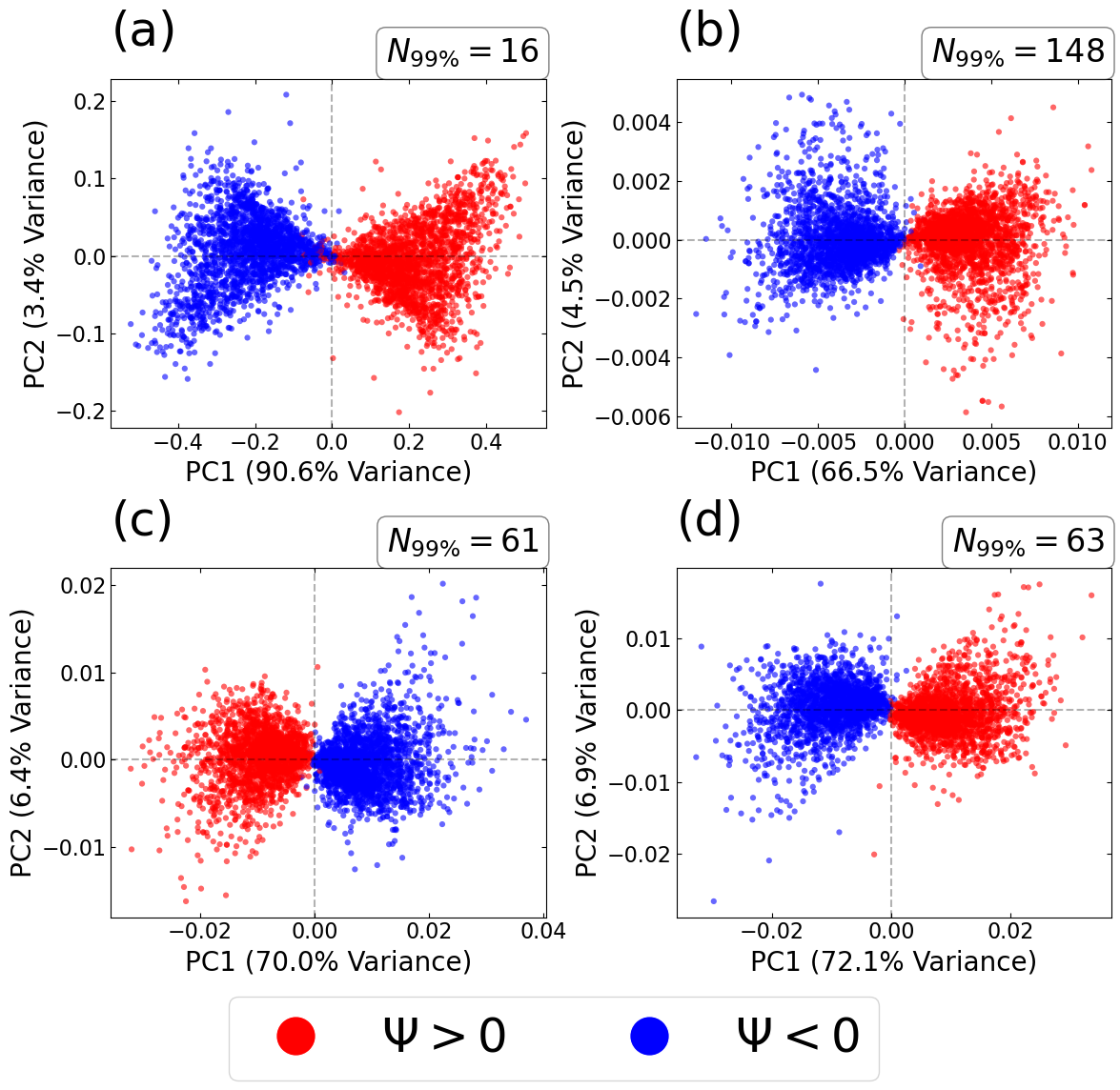}
    \centering
    \caption{PCA for the determinant basis. Projections of $4,096$ configurations onto the first two principal components (PC1, PC2) for architectures (a) Net 1, (b) Net 2, (c) Net 3, and (d) Net 4. Points are colored by the sign of the total wavefunction $\Psi$. The value $N_{99\%}$ indicates the number of principal components required to capture 99$\%$ of the total variance. 
    }
    \label{fig:pca_plots}
\end{figure}

Fig.~\ref{fig:pca_plots} presents the PCA results for the four network models. The projections illustrate a systematic refinement of the wavefunction's sign structure as the network architecture increases in complexity. In the baseline Net 1 (a), despite PC1 capturing $90.6\%$ of the variance, significant sign-mixing is observed near the origin, possibly indicating a poorly resolved nodal boundary. This model also exhibits the smallest $N_{99\%}$ value of only 16. Net 2 (b) demonstrates a shift toward a higher-dimensional basis representation, with $N_{99\%}$ increasing to $148$, yet the separation in the projection between the $\Psi > 0$ and $\Psi < 0$ domains remains diffuse. 

A qualitative transition occurs in the deep backflow architectures, Net 3 (c) and Net 4 (d), where the configuration clusters become clearly isolated along the PC1 axis. Notably, while these models achieve lower variational energies, their $N_{99\%}$ values drop significantly to around 60. This may indicate that the deep backflow mechanism represents the relevant sign variations using a smaller number of dominant principal components. With the larger embedding dimension, Net 4 (d) shows a more clearly defined boundary between the two clusters, with no visible sign-mixing. This model also exhibits a slight increase in the PC1 variance ($72.1\%$) and $N_{99\%}$ ($63$) compared to Net 3. Taken together, these results indicate that the deeper backflow architectures produce a more clearly separated representation of the wavefunction signs in the sampled configuration space. This separation, however, characterizes the learned representation rather than providing an independent test of whether the exact ground-state signs are correctly reproduced.

\section{Discussion}
In this work, we have systematically evaluated the performance of a deep neural network backflow transformation with low-rank multi-determinant updates for the variational representation of the 2D Hubbard model at $U/t=8$. We examined both the half-filled and 1/8-doped cases under periodic boundary conditions. By introducing a diagnostic framework based on Gram matrix analysis, singular-value spectra, and PCA of the determinant basis, we move beyond treating the neural network as a black box for energy minimization. Instead, these diagnostics provide an interpretable view of how different network architectures organize and utilize the multi-determinant basis, offering insight into the representation and efficiency of neural quantum states.

This diagnostic framework provides a two-stage characterization of the neural network's internal representation. The Gram matrix and singular-value spectrum characterize the pairwise similarity and effective dimensionality of the determinant basis, while PCA and the $N_{99\%}$ metric provide a complementary view of how efficiently the sampled configurations are represented in a lower-dimensional space. Our results reveal two directions for improving the basis representation of the neural wavefunction. First, the introduction of configuration-dependent weights enhances the diversity and effective dimensionality of the determinant expansion. By making the determinant coefficients $c_{ij}(\mathbf{x})$ functions of the input configurations, the network gains additional variational flexibility in combining the multi-determinant basis states. This is reflected in Nets 2, 3, and 4, which all achieve a threshold-dependent effective rank of approximately $280$--$289$.

Second, the deep backflow network significantly improves the representation of the wavefunction. While Net 2 increases basis diversity, its large $N_{99\%}$ value, together with the diffuse sign separation in the PCA projection, suggests that the resulting basis remains relatively distributed across many principal components and does not exhibit a clearly resolved sign boundary. In contrast, despite having a slightly lower effective rank and $N_{99\%}$ than Net 2, Net 3 and Net 4 achieve much lower variational energies and show a more distinct separation between the $\Psi>0$ and $\Psi<0$ configurations. This suggests that the deep backflow architectures can capture the relevant sign variations with a smaller number of dominant principal components while achieving lower variational energies. These results suggest that increasing the diversity of the determinant basis and improving the quality of its representation are distinct aspects of the neural wavefunction: configuration-dependent weights allow more determinants to contribute distinctively, while deep backflow enables these determinants to be combined in a way that more effectively captures the complex sign structure of the 2D Hubbard model.

Potential improvements to the multi-determinant ansatz could focus on increasing the expressivity and reducing redundancy of the basis. While low-rank matrix shifts reduce the number of parameters and accelerate determinant calculations (see Appendix~\ref{app:eff_det_cal}), employing full-rank matrices could provide greater flexibility in the determinant transformations, potentially leading to a more diverse basis at a higher computational cost. Another possible direction is to introduce an explicit regularization term that encourages reduced pairwise overlap between determinant basis states, applied as an auxiliary loss alongside energy minimization. Such a regularization could provide a direct way to control basis redundancy during training rather than relying solely on the energy objective.

\section{Acknowledgements}
Simulations for this work were performed using code built upon NetKet~\cite{netket3:2022,netket2:2019}. The neural network models and optimization pipelines were implemented with JAX \cite{jax2018github}, Flax \cite{flax2020github}, and Optax \cite{deepmind2020jax}. The implementation and training example are available in the accompanying GitHub repository~\cite{github_code}. We gratefully acknowledge the use of computational resources provided by the Yale Center for Research Computing.

\appendix
\section{Training and network hyperparameters}

\subsection{Training setup}
The parameter update $\delta \theta$ of the SPRING method at iteration $t$ is given by
\begin{equation}
    \delta \theta_t = \bar{O}_t^\top (\bar{O}_t \bar{O}_t^\top + \lambda I)^{-1} (\bar{\epsilon}_t - \mu \bar{O}_t \delta \theta_{t-1}) + \mu \delta \theta_{t-1},
\end{equation}
where $\bar{O}$ is the jacobian matrix with shape $(N_\text{sample}, N_\text{parameter})$
\begin{align}
    \begin{split}
        \bar{O}_{mk} = \frac{1}{\sqrt{N_{sample}}} (O_{mk} - \langle O_k \rangle)
    \end{split}
\end{align}
and 
\begin{equation}
    O_{m k} = \frac{\partial \ln \mathrm{\Psi}_\theta(\textbf{x}_m)}{\partial \theta_k} = \frac{1}{\mathrm{\Psi}_\theta(\textbf{x}_m)} \frac{\partial \mathrm{\Psi}_\theta(\textbf{x}_m)}{\partial \theta_k};
\end{equation}
$I$ is the identity matrix with shape $(N_\text{sample},N_\text{sample})$; \begin{equation}
        \bar{\epsilon}_m = \frac{-\delta \tau (E_L(\textbf{x}_m) - E)}{\sqrt{N_{sample}}}
\end{equation}
where $\delta \tau$ is the imaginary time step; $\lambda$ is the regularization diagonal shift for the ill-conditioned $\bar{O}$, and $\mu$ is a damping parameter used to stabilize the optimization process.

\begin{table}[h]
\centering
\caption{Parameter setup for the training process.}
\label{tab:VMC_and_SPRING}
\begin{tabular}{l c l}
\toprule
\textbf{Parameter} & \textbf{Symbol} & \textbf{Value} \\
\midrule
\multicolumn{3}{l}{\textit{VMC Sampling}} \\
Number of iteration & $N_{\text{step}}$ & 60000 \\
Number of samples & $N_{\text{sample}}$ & 4096 \\
Number of Markov chains & $N_{\text{chain}}$ & 16 \\
Sweep size & $N_\text{sweep}$ & $N_{\text{site}}$ \\
Discarded samples per chain & $N_{\text{discard}}$ & $N_{\text{site}}$ \\
\midrule
\multicolumn{3}{l}{\textit{SPRING Optimizer}} \\
Learning rate & $\eta$ & 0.01 \\
Diagonal shift & $\lambda$ & 0.001 \\
Momentum parameter & $\mu$ & 0.8 \\
Norm constraint & $C$ & $0.01 \to 0.01/9$\textsuperscript{*} \\
\bottomrule
\multicolumn{3}{l}{\textsuperscript{*}\footnotesize The value is adjusted after 50000 iterations.}
\end{tabular}
\end{table}

To ensure a stable optimization procedure, the SPRING algorithm introduces a norm constraint to the weight update. The constraint puts an additional lower bound on the constant learning rate $\eta$:
\begin{equation}
    \delta \theta = \delta \theta \cdot \mathrm{min}(\eta,\frac{C}{\|\delta w\|}),
\end{equation}
where $\|\delta w\|$ is the Euclidean norm of the weight updates, and $C$ is a scalar constraint.

Unless otherwise specified, the parameter configurations for the VMC and SPRING optimizer used in our experiments are listed in Table~\ref{tab:VMC_and_SPRING}.

\subsection{Network setup}
Table~\ref{tab:network_parameters} summarizes the default architectural hyperparameters for the neural network models employed throughout this study.

\begin{table}[h]
\centering
\caption{Hyperparameters for the network model.}
\label{tab:network_parameters}
\begin{tabular}{l c l}
\toprule
\textbf{Parameter} & \textbf{Symbol} & \textbf{Value} \\
\midrule
Embedding dimension & $d$ & 32 \\
Number of encoder layers & $N_\text{layer}$ & 3 \\
Number of static kernels & $N_\text{kernel}$ & 8 \\
Patch dimension & $p$ & 2 \\
Nonzero determinant shifts & $(N_{\Delta M}, N_{\Delta F})$ & (16, 16) \\
Low-rank cutoff & $r$ & 2 \\
\bottomrule
\end{tabular}
\end{table}

\subsection{Dimension of the matrices}
Table \ref{tab:appendix_matrix_dims} provides a comprehensive summary of the dimensions for all key matrices utilized throughout the neural network architecture.

\begin{table}[ht]
\centering
\setlength{\tabcolsep}{30pt} 
\caption{Dimensionality of key matrices.}
\label{tab:appendix_matrix_dims}
\begin{tabular}{ll}
\toprule
\textbf{Matrix} & \textbf{Dimension} \\
\midrule
$M$  & $2N_\text{site} \times N_\text{electron}$ \\
$F$ & $2N_\text{site} \times N_\text{electron}$ \\
$F'$ & $N_\text{site} \times N_\text{electron}$ \\
$F''$ & $2N_\text{site} \times N_\text{electron}$ \\
$J$ & $2N_\text{site} \times N_\text{electron}$ \\
$W_1$ & $2N_\text{site} \times N_\text{patch}$ \\
$Y_1$ & $d \times N_\text{electron}$ \\
$W_2$ & $2N_\text{site} \times N_\text{site}$ \\
$Y_2$ & $N_\text{electron} \times N_\text{electron}$ \\
$\Delta M_i$ & $2N_\text{site} \times N_\text{electron}$ \\
$\Delta F_j$ & $2N_\text{site} \times N_\text{electron}$ \\
$F_0$ & $2N_\text{site} \times N_\text{electron}$ \\
$U$ & $2N_\text{site} \times r$ \\
$V$ & $r \times N_\text{electron}$ \\
\bottomrule
\end{tabular}
\end{table}

\section{Derivation of matrix determinant lemma}
Consider the square block matrix $M \in \mathbb{R}^{(n+r)\times(n+r)}$ defined as
\begin{equation}
    M = 
    \begin{pmatrix}
        A & -U \\
        V^\top & I_r
    \end{pmatrix},
\end{equation}
where $A \in \mathbb{R}^{n \times n}$ is invertible, $U, V \in \mathbb{R}^{n \times r}$, and $I_r$ is the $r \times r$ identity matrix.

We can perform a block Gaussian elimination to factor $M$ as
\begin{equation}
    M =
    \begin{pmatrix}
        I_n & 0 \\
        V^\top A^{-1} & I_r
    \end{pmatrix}
    \begin{pmatrix}
        A & -U \\
        0 & I_r + V^\top A^{-1} U
    \end{pmatrix}.
\end{equation}
Both matrices on the right-hand side are block-triangular, so their determinants are given by the product of their diagonal blocks. Using multiplicativity of the determinant, we obtain
\begin{equation}
\det(M) = \det(A)\,\det\!\left(I_r + V^\top A^{-1} U\right).
\end{equation}

Alternatively, since $I_r$ is invertible, we may factor $M$ by pivoting on the lower-right block. Performing block elimination yields
\begin{equation}
    M =
    \begin{pmatrix}
        A + U V^\top & -U \\
        0 & I_r
    \end{pmatrix}
    \begin{pmatrix}
        I_n & 0 \\
        V^\top & I_r
    \end{pmatrix}.
\end{equation}
Taking determinants and using the block-triangular structure gives
\begin{equation}
\det(M) = \det(A + U V^\top).
\end{equation}

Equating the two expressions for $\det(M)$, we conclude
\begin{equation}
\det(A + U V^\top) = \det(A)\,\det\!\left(I_r + V^\top A^{-1} U\right),
\end{equation}
which proves the matrix determinant lemma.

\bibliography{bibifile}

@article{schollwock2005density,
  title={The density-matrix renormalization group},
  author={Schollw{\"o}ck, Ulrich},
  journal={Reviews of modern physics},
  volume={77},
  number={1},
  pages={259--315},
  year={2005},
  publisher={APS}
}

@article{stoudenmire2012studying,
  title={Studying two-dimensional systems with the density matrix renormalization group},
  author={Stoudenmire, Edwin M and White, Steven R},
  journal={Annu. Rev. Condens. Matter Phys.},
  volume={3},
  number={1},
  pages={111--128},
  year={2012},
  publisher={Annual Reviews}
}

@article{vidal2007classical,
  title={Classical simulation of infinite-size quantum lattice systems in one spatial dimension},
  author={Vidal, Guifr{\'e}},
  journal={Physical review letters},
  volume={98},
  number={7},
  pages={070201},
  year={2007},
  publisher={APS}
}

@article{cirac2021matrix,
  title={Matrix product states and projected entangled pair states: Concepts, symmetries, theorems},
  author={Cirac, J Ignacio and Perez-Garcia, David and Schuch, Norbert and Verstraete, Frank},
  journal={Reviews of Modern Physics},
  volume={93},
  number={4},
  pages={045003},
  year={2021},
  publisher={APS}
}

@article{kraus2010fermionic,
  title={Fermionic projected entangled pair states},
  author={Kraus, Christina V and Schuch, Norbert and Verstraete, Frank and Cirac, J Ignacio},
  journal={Physical Review A—Atomic, Molecular, and Optical Physics},
  volume={81},
  number={5},
  pages={052338},
  year={2010},
  publisher={APS}
}

@article{zhang1997constrained,
  title={Constrained path Monte Carlo method for fermion ground states},
  author={Zhang, Shiwei and Carlson, Joseph and Gubernatis, James E},
  journal={Physical Review B},
  volume={55},
  number={12},
  pages={7464},
  year={1997},
  publisher={APS}
}

@article{zhang2003quantum,
  title={Quantum Monte Carlo method using phase-free random walks with Slater determinants},
  author={Zhang, Shiwei and Krakauer, Henry},
  journal={Physical review letters},
  volume={90},
  number={13},
  pages={136401},
  year={2003},
  publisher={APS}
}

@article{sugiyama1986auxiliary,
  title={Auxiliary field Monte-Carlo for quantum many-body ground states},
  author={Sugiyama, G and Koonin, SE},
  journal={Annals of Physics},
  volume={168},
  number={1},
  pages={1--26},
  year={1986},
  publisher={Elsevier}
}

@article{zhang201315,
  title={15 Auxiliary-Field Quantum Monte Carlo for Correlated Electron Systems},
  author={Zhang, Shiwei},
  journal={Emergent Phenomena in Correlated Matter},
  year={2013}
}

@book{becca2017quantum,
  title={Quantum Monte Carlo approaches for correlated systems},
  author={Becca, Federico and Sorella, Sandro},
  year={2017},
  publisher={Cambridge University Press}
}

@article{carleo2017solving,
  title={Solving the quantum many-body problem with artificial neural networks},
  author={Carleo, Giuseppe and Troyer, Matthias},
  journal={Science},
  volume={355},
  number={6325},
  pages={602--606},
  year={2017},
  publisher={American Association for the Advancement of Science}
}

@article{wu2024variational,
  title={Variational benchmarks for quantum many-body problems},
  author={Wu, Dian and Rossi, Riccardo and Vicentini, Filippo and Astrakhantsev, Nikita and Becca, Federico and Cao, Xiaodong and Carrasquilla, Juan and Ferrari, Francesco and Georges, Antoine and Hibat-Allah, Mohamed and others},
  journal={Science},
  volume={386},
  number={6719},
  pages={296--301},
  year={2024},
  publisher={American Association for the Advancement of Science}
}

@article{schmitt2022quantum,
  title={Quantum phase transition dynamics in the two-dimensional transverse-field Ising model},
  author={Schmitt, Markus and Rams, Marek M and Dziarmaga, Jacek and Heyl, Markus and Zurek, Wojciech H},
  journal={Science Advances},
  volume={8},
  number={37},
  pages={eabl6850},
  year={2022},
  publisher={American Association for the Advancement of Science}
}

@article{nys2024ab,
  title={Ab-initio variational wave functions for the time-dependent many-electron Schr{\"o}dinger equation},
  author={Nys, Jannes and Pescia, Gabriel and Sinibaldi, Alessandro and Carleo, Giuseppe},
  journal={Nature communications},
  volume={15},
  number={1},
  pages={9404},
  year={2024},
  publisher={Nature Publishing Group UK London}
}

@article{chen2024empowering,
  title={Empowering deep neural quantum states through efficient optimization},
  author={Chen, Ao and Heyl, Markus},
  journal={Nature Physics},
  volume={20},
  number={9},
  pages={1476--1481},
  year={2024},
  publisher={Nature Publishing Group UK London}
}

@article{rende2024simple,
  title={A simple linear algebra identity to optimize large-scale neural network quantum states},
  author={Rende, Riccardo and Viteritti, Luciano Loris and Bardone, Lorenzo and Becca, Federico and Goldt, Sebastian},
  journal={Communications Physics},
  volume={7},
  number={1},
  pages={260},
  year={2024},
  publisher={Nature Publishing Group UK London}
}

@article{ou2025improving,
  title={Improving neural network performance for solving quantum sign structure},
  author={Ou, Xiaowei and Huang, Tianshu and Ozoli{\c{n}}{\v{s}}, Vidvuds},
  journal={Physical Review B},
  volume={112},
  number={16},
  pages={165122},
  year={2025},
  publisher={APS}
}

@article{viteritti2025transformer,
  title={Transformer wave function for two dimensional frustrated magnets: Emergence of a spin-liquid phase in the Shastry-Sutherland model},
  author={Viteritti, Luciano Loris and Rende, Riccardo and Parola, Alberto and Goldt, Sebastian and Becca, Federico},
  journal={Physical Review B},
  volume={111},
  number={13},
  pages={134411},
  year={2025},
  publisher={APS}
}

@article{nomura2017restricted,
  title={Restricted Boltzmann machine learning for solving strongly correlated quantum systems},
  author={Nomura, Yusuke and Darmawan, Andrew S and Yamaji, Youhei and Imada, Masatoshi},
  journal={Physical Review B},
  volume={96},
  number={20},
  pages={205152},
  year={2017},
  publisher={APS}
}

@article{choo2020fermionic,
  title={Fermionic neural-network states for ab-initio electronic structure},
  author={Choo, Kenny and Mezzacapo, Antonio and Carleo, Giuseppe},
  journal={Nature communications},
  volume={11},
  number={1},
  pages={2368},
  year={2020},
  publisher={Nature Publishing Group UK London}
}

@article{pfau2020ab,
  title={Ab initio solution of the many-electron Schr{\"o}dinger equation with deep neural networks},
  author={Pfau, David and Spencer, James S and Matthews, Alexander GDG and Foulkes, W Matthew C},
  journal={Physical review research},
  volume={2},
  number={3},
  pages={033429},
  year={2020},
  publisher={APS}
}

@article{hermann2020deep,
  title={Deep-neural-network solution of the electronic Schr{\"o}dinger equation},
  author={Hermann, Jan and Sch{\"a}tzle, Zeno and No{\'e}, Frank},
  journal={Nature Chemistry},
  volume={12},
  number={10},
  pages={891--897},
  year={2020},
  publisher={Nature Publishing Group UK London}
}

@article{liu2024neural,
  title={Neural network backflow for ab initio quantum chemistry},
  author={Liu, An-Jun and Clark, Bryan K},
  journal={Physical Review B},
  volume={110},
  number={11},
  pages={115137},
  year={2024},
  publisher={APS}
}

@article{luo2019backflow,
  title={Backflow transformations via neural networks for quantum many-body wave functions},
  author={Luo, Di and Clark, Bryan K},
  journal={Physical review letters},
  volume={122},
  number={22},
  pages={226401},
  year={2019},
  publisher={APS}
}

@article{robledo2022fermionic,
  title={Fermionic wave functions from neural-network constrained hidden states},
  author={Robledo Moreno, Javier and Carleo, Giuseppe and Georges, Antoine and Stokes, James},
  journal={Proceedings of the National Academy of Sciences},
  volume={119},
  number={32},
  pages={e2122059119},
  year={2022},
  publisher={National Academy of Sciences}
}

@article{liu2024unifying,
  title={Unifying view of fermionic neural network quantum states: From neural network backflow to hidden fermion determinant states},
  author={Liu, Zejun and Clark, Bryan K},
  journal={Physical Review B},
  volume={110},
  number={11},
  pages={115124},
  year={2024},
  publisher={APS}
}

@article{chen2025neural,
  title={Neural network-augmented pfaffian wave-functions for scalable simulations of interacting fermions},
  author={Chen, Ao and Wan, Zhou-Quan and Sengupta, Anirvan and Georges, Antoine and Roth, Christopher},
  journal={arXiv preprint arXiv:2507.10705},
  year={2025}
}

@article{gu2025solving,
  title={Solving the hubbard model with neural quantum states},
  author={Gu, Yuntian and Li, Wenrui and Lin, Heng and Zhan, Bo and Li, Ruichen and Huang, Yifei and He, Di and Wu, Yantao and Xiang, Tao and Qin, Mingpu and others},
  journal={arXiv preprint arXiv:2507.02644},
  year={2025}
}

@article{goldshlager2024kaczmarz,
  title={A Kaczmarz-inspired approach to accelerate the optimization of neural network wavefunctions},
  author={Goldshlager, Gil and Abrahamsen, Nilin and Lin, Lin},
  journal={Journal of Computational Physics},
  volume={516},
  pages={113351},
  year={2024},
  publisher={Elsevier}
}

@article{sorella2005wave,
  title={Wave function optimization in the variational Monte Carlo method},
  author={Sorella, Sandro},
  journal={Physical Review B—Condensed Matter and Materials Physics},
  volume={71},
  number={24},
  pages={241103},
  year={2005},
  publisher={APS}
}

@article{sharma2025comparing,
  title={Comparing Symmetrized Determinant Neural Quantum States for the Hubbard Model},
  author={Sharma, Louis and Shokry, Ahmedeo and Nutakki, Rajah and Simard, Olivier and Ferrero, Michel and Vicentini, Filippo},
  journal={arXiv preprint arXiv:2510.11710},
  year={2025}
}

@article{loehr2025enhancing,
  title={Enhancing Neural Network Backflow},
  author={Loehr, Kieran and Clark, Bryan K},
  journal={arXiv preprint arXiv:2510.26906},
  year={2025}
}

@article{romero2025spectroscopy,
  title={Spectroscopy of two-dimensional interacting lattice electrons using symmetry-aware neural backflow transformations},
  author={Romero, Imelda and Nys, Jannes and Carleo, Giuseppe},
  journal={Communications Physics},
  volume={8},
  number={1},
  pages={46},
  year={2025},
  publisher={Nature Publishing Group UK London}
}

@article{rodriguez2013multireference,
  title={Multireference symmetry-projected variational approaches for ground and excited states of the one-dimensional Hubbard model},
  author={Rodr{\'\i}guez-Guzm{\'a}n, R and Jim{\'e}nez-Hoyos, Carlos A and Schutski, R and Scuseria, Gustavo E},
  journal={Physical Review B—Condensed Matter and Materials Physics},
  volume={87},
  number={23},
  pages={235129},
  year={2013},
  publisher={APS}
}

@article{nomura2021helping,
  title={Helping restricted Boltzmann machines with quantum-state representation by restoring symmetry},
  author={Nomura, Yusuke},
  journal={Journal of Physics: Condensed Matter},
  volume={33},
  number={17},
  pages={174003},
  year={2021},
  publisher={IOP Publishing}
}

@article{reh2023optimizing,
  title={Optimizing design choices for neural quantum states},
  author={Reh, Moritz and Schmitt, Markus and G{\"a}rttner, Martin},
  journal={Physical Review B},
  volume={107},
  number={19},
  pages={195115},
  year={2023},
  publisher={APS}
}

@article{ding2007eigenvalues,
  title={Eigenvalues of rank-one updated matrices with some applications},
  author={Ding, Jiu and Zhou, Aihui},
  journal={Applied Mathematics Letters},
  volume={20},
  number={12},
  pages={1223--1226},
  year={2007},
  publisher={Elsevier}
}

@article{dosovitskiy2020image,
  title={An image is worth 16x16 words: Transformers for image recognition at scale},
  author={Dosovitskiy, Alexey and Beyer, Lucas and Kolesnikov, Alexander and Weissenborn, Dirk and Zhai, Xiaohua and Unterthiner, Thomas and Dehghani, Mostafa and Minderer, Matthias and Heigold, Georg and Gelly, Sylvain and others},
  journal={arXiv preprint arXiv:2010.11929},
  year={2020}
}

@article{feynman1956energy,
  title={Energy spectrum of the excitations in liquid helium},
  author={Feynman, Richard P and Cohen, Michael},
  journal={Physical Review},
  volume={102},
  number={5},
  pages={1189},
  year={1956},
  publisher={APS}
}

@article{hendrycks2016gaussian,
  title={Gaussian error linear units (gelus)},
  author={Hendrycks, Dan and Gimpel, Kevin},
  journal={arXiv preprint arXiv:1606.08415},
  year={2016}
}

@article{qin2016benchmark,
  title={Benchmark study of the two-dimensional Hubbard model with auxiliary-field quantum Monte Carlo method},
  author={Qin, Mingpu and Shi, Hao and Zhang, Shiwei},
  journal={Physical Review B},
  volume={94},
  number={8},
  pages={085103},
  year={2016},
  publisher={APS}
}

@article{zhou2024solving,
  title={Solving Fermi-Hubbard-type models by tensor representations of backflow corrections},
  author={Zhou, Yu-Tong and Zhou, Zheng-Wei and Liang, Xiao},
  journal={Physical Review B},
  volume={109},
  number={24},
  pages={245107},
  year={2024},
  publisher={APS}
}

@Article{netket3:2022,
    title={NetKet 3: Machine Learning Toolbox for Many-Body Quantum Systems},
    author={Filippo Vicentini and Damian Hofmann and Attila Szabó and Dian Wu and Christopher Roth and Clemens Giuliani and Gabriel Pescia and Jannes Nys and Vladimir Vargas-Calderón and Nikita Astrakhantsev and Giuseppe Carleo},
    journal={SciPost Phys. Codebases},
    pages={7},
    year={2022},
    publisher={SciPost},
    doi={10.21468/SciPostPhysCodeb.7},
    url={https://scipost.org/10.21468/SciPostPhysCodeb.7}
}

@article{netket2:2019,
    title={NetKet: A machine learning toolkit for many-body quantum systems},
    author={Carleo, Giuseppe and Choo, Kenny and Hofmann, Damian and Smith, James ET and Westerhout, Tom and Alet, Fabien and Davis, Emily J and Efthymiou, Stavros and Glasser, Ivan and Lin, Sheng-Hsuan and Mauri, Marta and Mazzola, Guglielmo and Pereira, Christian B and Vicentini, Filippo},
    journal={SoftwareX},
    volume={10},
    pages={100311},
    year={2019},
    publisher={Elsevier},
    doi={10.1016/j.softx.2019.100311},
    url={https://www.sciencedirect.com/science/article/pii/S2352711019300974}
}

@software{jax2018github,
  author = {James Bradbury and Roy Frostig and Peter Hawkins and Matthew James Johnson and Chris Leary and Dougal Maclaurin and George Necula and Adam Paszke and Jake Vander{P}las and Skye Wanderman-{M}ilne and Qiao Zhang},
  title = {{JAX}: composable transformations of {P}ython+{N}um{P}y programs},
  url = {http://github.com/jax-ml/jax},
  version = {0.3.13},
  year = {2018},
}

@software{flax2020github,
  author = {Jonathan Heek and Anselm Levskaya and Avital Oliver and Marvin Ritter and Bertrand Rondepierre and Andreas Steiner and Marc van {Z}ee},
  title = {{F}lax: A neural network library and ecosystem for {JAX}},
  url = {http://github.com/google/flax},
  version = {0.12.6},
  year = {2024},
}

@software{deepmind2020jax,
  title = {The {D}eep{M}ind {JAX} {E}cosystem},
  author = {DeepMind and Babuschkin, Igor and Baumli, Kate and Bell, Alison and Bhupatiraju, Surya and Bruce, Jake and Buchlovsky, Peter and Budden, David and Cai, Trevor and Clark, Aidan and Danihelka, Ivo and Dedieu, Antoine and Fantacci, Claudio and Godwin, Jonathan and Jones, Chris and Hemsley, Ross and Hennigan, Tom and Hessel, Matteo and Hou, Shaobo and Kapturowski, Steven and Keck, Thomas and Kemaev, Iurii and King, Michael and Kunesch, Markus and Martens, Lena and Merzic, Hamza and Mikulik, Vladimir and Norman, Tamara and Papamakarios, George and Quan, John and Ring, Roman and Ruiz, Francisco and Sanchez, Alvaro and Sartran, Laurent and Schneider, Rosalia and Sezener, Eren and Spencer, Stephen and Srinivasan, Srivatsan and Stanojevi\'{c}, Milo\v{s} and Stokowiec, Wojciech and Wang, Luyu and Zhou, Guangyao and Viola, Fabio},
  url = {http://github.com/google-deepmind},
  year = {2020},
}

@misc{github_code,
  howpublished = {\url{https://github.com/sweetpotat5/Multi-determinant-nerual-network-backflow}}
}

\end{document}